\documentclass[twocolumn,showpacs,preprintnumbers,amsmath,amssymb]{revtex4}
\usepackage[pdftex]{epsfig}
\usepackage{dcolumn}
\usepackage{bm}
\DeclareGraphicsExtensions{.jpg,.mps,.pdf,.pdf,.png}

\newcommand{\beq}{\begin{equation}} 
\newcommand{\eeq}{\end{equation}} 
\newcommand{\beqa}{\begin{eqnarray}} 
\newcommand{\eeqa}{\end{eqnarray}} 
\newcommand{\bea}{\begin{array}} 
\newcommand{\ea}{\end{array}} 

\newcommand{\dd}{{\rm d}}
\renewcommand{\pl}{\partial}
 
\newcommand{\lag}{\langle} 
\newcommand{\rag}{\rangle} 
\newcommand{\law}{\stackrel{\rm law}{=}}
\newcommand{\ii}{{\rm i}}
\newcommand{\llag}{\prec\!} 
\newcommand{\rrag}{\!\succ}

\newcommand{\bu}{{\bf u}}
\newcommand{\bx}{{\bf x}}
\newcommand{\bk}{{\bf k}}
\newcommand{\bq}{{\bf q}}
\newcommand{\bv}{{\bf v}}

\newcommand{\vx}{{\bf x}}

\newcommand{\vq}{{\bf q}}

\newcommand{\thetat}{{\tilde{\theta}}}

\newcommand{\cP}{{\cal P}}
\newcommand{\cL}{{\cal L}}
\newcommand{\bs}{{\bf s}}
\newcommand{\cV}{{\cal V}}
\newcommand{\bQ}{{\bf Q}}
\newcommand{\bX}{{\bf X}}
\newcommand{\bU}{{\bf U}}

\newcommand{\ut}{{\tilde{u}}}
\newcommand{\cC}{{\cal C}}

\newcommand{\bP}{{\bf P}}
\newcommand{\mS}{{\cal S}}
\newcommand{\bp}{{\bf p}}

\newcommand{\cA}{{\cal A}}
\newcommand{\cB}{{\cal B}}
\newcommand{\cD}{{\cal D}}

\begin{document}

\title{Merging and fragmentation in the Burgers dynamics}

\author{Francis Bernardeau}
\affiliation{Institut de Physique Th{\'e}orique,
    CEA/DSM/IPhT, Unit{\'e} de recherche associ{\'e}e au CNRS, CEA/Saclay,
    91191 Gif-sur-Yvette c{\'e}dex, France}
\author{Patrick Valageas}
\affiliation{Institut de Physique Th{\'e}orique,
    CEA/DSM/IPhT, Unit{\'e} de recherche associ{\'e}e au CNRS, CEA/Saclay,
    91191 Gif-sur-Yvette c{\'e}dex, France}
\vspace{.2 cm}

\date{\today}
\vspace{.2 cm}

\begin{abstract}

We explore the noiseless Burgers dynamics in the inviscid limit, the so-called ``adhesion
model'' in cosmology, in a regime where (almost) all the fluid particles are embedded within
point-like massive halos. Following previous works, we focus our investigations on a 
``geometrical'' model, where the matter evolution within the shock manifold is defined from
a geometrical construction. This hypothesis is at variance with the assumption that the usual
continuity equation holds but, in the inviscid limit, both models agree in the regular regions. 

Taking advantage of the formulation of the dynamics of this  ``geometrical model''
in terms of Legendre transforms and convex hulls,
we study the evolution with time of the distribution of matter and the associated
partitions of the Lagrangian and Eulerian spaces. We describe how the halo mass
distribution derives from a triangulation in Lagrangian space, while the dual
Voronoi-like tessellation in Eulerian space gives the boundaries of empty regions
with shock nodes at their vertices.

We then emphasize that this dynamics actually leads to halo
fragmentations for space dimensions greater or equal to 2 (for the inviscid limit studied
in this article). This is most easily
seen from the properties of the Lagrangian-space triangulation and we illustrate this
process in the two-dimensional (2D) case. In particular, we explain how point-like halos only
merge through three-body collisions while two-body collisions always give rise to two new
massive shock nodes (in 2D). This generalizes to higher dimensions and we briefly
illustrate the three-dimensional (3D) case. This leads to a specific picture for the continuous
formation of massive halos through successive halo fragmentations and mergings.

\end{abstract}

 \vskip2pc

\maketitle

\section{Introduction}
\label{sec:intro}

The Burgers equation, introduced at the end of the 1930s in \cite{Burgers1939}, was
proposed as an archetype system for the dynamics of pressureless gases. It indeed shares
with the Navier-Stokes equations its quadratic nonlinearity, and its symmetry properties
(invariance under parity, under space and time translations). It fails however to reproduce
the chaotic behaviors one encounters in Navier-Stokes systems, as it became clear when
an explicit fully deterministic solution of this system was found by Hopf \cite{Hopf1950}
and Cole \cite{Cole1951}. Nevertheless, the Burgers dynamics has retained much interest
for hydrodynamical studies, as a benchmark for approximation schemes \cite{Fournier1983}
and as a simpler example of intermittency phenomena \cite{Kida1979,She1992,Sinai1992}.
Moreover, the Burgers equation has also appeared in many physical contexts,
such as the propagation of nonlinear acoustic waves in
non-dispersive media \cite{Gurbatov1991} or the formation of large-scale
structures in cosmology \cite{Gurbatov1989,Vergassola1994},
see \cite{Bec2007} for a recent review.

In particular, in the cosmological context it provides an approximate description of the
nonlinear evolution of the large-scale structure of the Universe, and it leads to
complex structures (similar to the network of filaments observed between clusters
of galaxies in simulations or in the sky) in dimension two or larger
\cite{1990MNRAS.242..200K,Weinberg1990,1992ApJ...393..437K,Vergassola1994,Melott1994,Saichev1997}.
In this cosmological context (and also in many hydrodynamical studies) one is interested
in the noiseless inviscid limit, where the viscosity is sent to zero and when the initial velocity
field obeys Gaussian statistics. Then, regular fluid elements are
simply free moving, until singularities (shocks) appear. There, the infinitesimal
viscosity prevents matter flows from crossing each other and gives rise to shocks and
massive local structures of various co-dimensions, of infinitesimal width.
This effect mimics the formation of gravitationally bound objects
and allows to reproduce the skeleton of the large scale structure built by the exact
gravitational dynamics with the same initial conditions 
\cite{1992ApJ...393..437K,Weinberg1990}. Then, this model is often referred to as the
``adhesion model'', following Ref.~\cite{Gurbatov1989}.

Many works have been devoted to the early stages of the dynamics (see \cite{FrischBec2001}
and references in \cite{Bec2007}) of such models, where only a fraction of the mass has
reached caustics, and to
the universal exponents associated with the formation of these singularities. 
In this paper our interest is more particularly focused
on a limit where
almost all the fluid particles have reached caustics and are actually in point-like massive
objects. This is the regime of interest for cosmological purposes, where the
stochastic initial velocity field is Gaussian with a power-law (or slowly running) power
spectrum (analogous to fractional Brownian motion in 1D).
Moreover, we are mostly interested in the distribution of matter (i.e. the density field
generated by the Burgers velocity field, starting with a uniform initial density),
rather than in the properties of the velocity field itself.

Here we must note that the Burgers equation itself, and its Hopf-Cole solution, refers
to the equation of motion for the velocity field. Since we
are interested in the evolution of the matter distribution, we need to couple this velocity
field to a density field and follow their simultaneous evolution. Note that a significant
difference with the gravitational dynamics is that the evolution of the velocity field is
independent from that of the density field, which in this sense plays a passive role.
The ``standard'' procedure would be to use the usual continuity equation for the
density field, and next take the inviscid limit $\nu\rightarrow 0^+$.
This is a natural and well-defined model which has been studied for instance
in Refs.~\cite{Bogaevsky2004,Bogaevsky2006}.
A drawback of this procedure is that, analytically, it only provides
information on the instantaneous Eulerian velocity field inside
shocks; to fully obtain the Lagrangian map, numerical integration
cannot be avoided.
One then loses the whole interest of having the Hopf-Cole solution at our disposal.

There exists however an alternative procedure for the evolution of the matter distribution
which is embedded in the Hopf-Cole solution for the velocity field.
%
Indeed, as noticed in Refs.~\cite{Gurbatov1991,Saichev1996,Saichev1997}, in the inviscid limit 
the Hopf-Cole solution implicitly provides an ``inverse Lagrangian map'', $\vx\mapsto \vq$, which for regular points gives the
Lagrangian coordinate $\vq$ of the particle that is located at position $\vx$ at time $t$.
Once shocks have formed, this map is not regular anymore and cannot be inverted 
without ambiguities (except in the one-dimensional case), since the
``interior'' of shock nodes are not reached when the whole Eulerian space is spanned. 
However, using the fact that the ``inverse Lagrangian map'',
$\vx\mapsto \vq$, can be written in terms of a Legendre transform, it is possible to
{\it define} the ``direct map'', $\vq\mapsto\vx$, by Legendre conjugacy.
In fact, this mapping can also be built as the inviscid limit of a specific mapping
$\vx\leftrightarrow \vq$, which holds at finite $\nu$ where all points are regular. 
In this manner, to quote Ref.~\cite{Saichev1996}, one obtains an ``analytically convenient''
model for the density field, since the matter distribution can be obtained at any time
$t$ through Legendre transforms, or equivalent geometrical constructions, without the
need to explicitly solve a differential equation over all previous times.
Of course, this implies in turn that the density field defined by this ``geometrical model''
does not obey the standard continuity equation. More precisely, there appears a specific
diffusive-type term in the right hand side of the continuity equation. Then, in the
inviscid limit one recovers the standard behavior outside of shocks, but there remains
significant differences within the shock manifold. This is particularly manifest when one tries to follow 
the history of accretion/merging of mass clusters. 
As already noticed in Ref.~\cite{Gurbatov1991} (chapter 6, see also note\footnote{This property was also noticed by 
D. Pogosyan in a 1989 preprint which is unfortunately unpublished.}), in the geometrical model, and
for dimensions two and higher, mass clusters do not necessarily
merge when they collide, and the collision can give rise to several new clusters (with an
exchange of matter and a possibly different number of outgoing clusters, as we shall
explain in this paper). This is clearly qualitatively at variance with what the standard model gives rise to. 
In the latter case the resulting mass evolution follows indeed that of a 
genuine ``adhesion model'', although some unexpected behaviors are still encountered. For instance
mass clusters can leave shock nodes and travel along the shock manifold
\cite{Bogaevsky2004,Bogaevsky2006}.

The standard model and the geometrical model clearly differ and while the first is based on a 
somewhat natural assumption regarding the continuity equation, the second  allows easier numerical and
analytical insights. Investigations of the latter model is further justified from numerical 
works that have shown that the latter provides a valuable model in the cosmological context, as it builds
large-scale structure and even shock mass functions that are similar to those
obtained in the standard gravitational dynamics \cite{Vergassola1994}. In any case, it provides a rare example of a
non-trivial mass transportation, coupled to a dynamical velocity field, which can be
integrated and where significant analytical results can be obtained.

In this paper we revisit the ``geometrical model'', focusing on the late time mass distribution 
after (almost) all the mass has reached the shock manifold and gathered in point-like objects. 
The Hopf-Cole solution, and its 
geometrical interpretation, provides a way to describe the mass distribution in this limit and it 
corresponds, in Eulerian space, to a ``Voronoi-like'' tessellation (as it was already noticed in
\cite{1992ApJ...393..437K} and further described in more details in \cite{Saichev1996,Saichev1997}, 
and in the review paper \cite{Woyczynski2007}).
We further explore the consequences of this construction in the context of cosmological studies
by paying special attention to the evolution with
time of the dual Lagrangian-space triangulation and Eulerian-space tessellation.
Thus, we describe how the halo masses can be obtained, by means of Legendre transforms
or geometrical constructions, and eventually how they are
rearranged as time evolves. In particular, this allows us to obtain the peculiar collision
rules that drive the dynamics and to illustrate with further details the exchanges of matter
that can take place during these events.
In this paper we cover those aspects in a rather qualitative way, as the focus will
be on the description of these mechanisms and their illustration with numerical
simulations in the two-dimensional case.
The outline of the paper is the following. The Burgers equation and the Hopf-Cole solution
are introduced in Sect.~\ref{Burgers-dynamics}. Its dual description in both Eulerian and 
Lagrangian spaces is presented in Sect.~\ref{Lagrangian-potential}, together with a precise
definition of the inviscid limit associated with the ``geometrical model'' and the resulting
late time behavior of the displacement
fields and potentials. Finally, numerical results are presented in Sect.~\ref{One-dimension}
and \ref{Two-dimensions} for respectively the $d=1$ and $d=2$ cases, as well as a
detailed description of the fragmentation-merging processes.

\section{Burgers dynamics}
\label{Burgers-dynamics}

\subsection{Equations of motion}
\label{eq-motion}

We consider the $d$-dimensional Burgers  equation (with $d\geq 1$),
\beq
\pl_t \bu + (\bu\cdot\nabla)\bu = \nu \Delta \bu,
\label{Burgers}
\eeq
in the inviscid limit, $\nu\rightarrow 0^+$,
for the velocity field $\bu(\bx,t)$, and the evolution of the density field 
$\rho(\bx,t)$ generated by this dynamics, starting from a uniform density $\rho_0$
at the initial time $t=0$. As pointed out in the introduction, in this article we study
a specific ``geometrical model'' for the distribution of matter, where the density
field is not coupled to the velocity field (\ref{Burgers}) through the standard continuity
equation but through the modified equation (\ref{cont1}) below. We shall discuss this 
``geometrical model'' in Sect.~\ref{Lagrangian-potential} and explain in details
its relationship with the Burgers dynamics (\ref{Burgers}). However, we must first
describe in this section the initial conditions that we consider in this article and the
properties of the Burgers velocity field.

It is a well known result that if the initial velocity is potential,
$\bu_0=-\nabla\psi_0$, 
it remains so forever \cite{Bec2007}, so that the velocity field is fully defined by its 
potential $\psi(\bx,t)$, or by its divergence $\theta(\bx,t)$, through
\beq
\bu=-\nabla\psi, \;\;\;\; \theta = -\nabla\cdot\bu = \Delta \psi .
\label{thetadef}
\eeq

Having in mind the use of the Burgers dynamics as a model for the formation of
the large-scale structure in cosmology, we shall consider for numerical implementations 
the case where $\psi_{0}$ is a random Gaussian field, statistically homogeneous and isotropic, 
thus entirely characterized by its power spectrum (see for instance \cite{Bernardeau2002} 
for details). The power spectrum 
of $\psi_{0}$ can be equivalently defined by that of the velocity divergence,
$P_{\theta_0}(k)$, such that,
\beq
\lag\thetat_0\rag=0 , \;\;\; \lag\thetat_0(\bk_1)\thetat_0(\bk_2)\rag = 
\delta_D(\bk_1+\bk_2) P_{\theta_0}(k_1) ,
\label{Ptheta0def}
\eeq
where we note $\delta_D$ the Dirac distribution and $\thetat_0$ the Fourier transform
of the initial divergence,
\beq
\theta_0(\bx) = \int\dd\bk \; e^{\ii\bk.\bx} \; \thetat_0(\bk).
\label{Fourier}
\eeq
For simplicity we shall also assume that the power spectrum $P_{\theta_0}(k)$ is a power
law of index $n$,
\beq
P_{\theta_0}(k) = \frac{D}{(2\pi)^d} \, k^{n+3-d} \;\;\; \mbox{with} \;\; -3<n<1 .
\label{ndef}
\eeq
As a result, for the range $-3<n<1$ the dynamics is self-similar: a rescaling of time is
statistically equivalent to a rescaling of distances, as
\beq
\lambda>0: \;\; t\rightarrow \lambda t, \;\; \bx \rightarrow 
\lambda^{2/(n+3)} \bx ,
\label{selfsimilar}
\eeq
so that at a given time $t$ there is only one characteristic scale, which we normalize as
\beq
L(t) =  (2Dt^2)^{1/(n+3)} .
\label{Lt}
\eeq
It marks the transition from the large-scale linear regime to the small-scale nonlinear
regime.
In order to express the scaling law (\ref{selfsimilar}) it is convenient to
introduce the dimensionless scaling variables
\beq
\bQ= \frac{\bq}{L(t)} , \;\;\; \bX= \frac{\bx}{L(t)} , \;\;\; \bU= \frac{t\bu}{L(t)} .
\label{QXU}
\eeq
Then, equal-time statistical quantities (such as correlations or probability distributions)
written in terms of these variables no longer depend on time and the scale $X=1$
is the characteristic scale of the system.

We stress however that many properties that we discuss in the following also apply to
more general initial conditions than (\ref{ndef}), such as those with a scale dependent power
spectrum with a local slope $n=d-3+\dd\ln P_{\theta_0}/\dd\ln k$ that could vary
with $k$ but remains in the range $]-3,1[$ at very small and very large scales.
This is typically what is expected in cosmology.

\subsection{Hopf-Cole solution}
\label{Hopf-Cole}

With the Hopf-Cole transformation  \cite{Hopf1950,Cole1951}, 
$\psi(\bx,t)=2\nu\ln\Xi(\bx,t)$, the nonlinear Burgers equation (\ref{Burgers}) transforms
into a linear heat equation for $\Xi(\bx,t)$, which leads to the solution
\beq
\psi(\bx,t) = 2\nu\ln\int\frac{\dd\bq}{(4\pi\nu t)^{d/2}} \, 
\exp\left[ \frac{\psi_0(\bq)}{2\nu}-\frac{|\bx-\bq|^2}{4\nu t}\right] .
\label{Hopf1}
\eeq
Then, in the inviscid limit $\nu\rightarrow 0^+$, a steepest-descent method
gives \cite{Burgersbook,Bec2007}
\beq
\psi(\bx,t) = \max_{\bq}\left[\psi_0(\bq)-\frac{|\bx-\bq|^2}{2t}\right] .
\label{psixpsi0q}
\eeq
If the maximum in (\ref{psixpsi0q}) is reached at a unique
point, $\bq(\bx,t)$, no shock has formed there and the quantity $\bq(\bx,t)$
is the Lagrangian coordinate of the particle that is
located at the Eulerian position $\bx$ at time $t$ \cite{Burgersbook,Bec2007}
(hereafter, we note by the letter $\bq$ the Lagrangian coordinates, i.e. the initial
positions at $t=0$ of particles, and by the letter $\bx$ the Eulerian coordinates
at any time $t>0$).
Moreover, still in the absence of shocks, this particle has kept its initial velocity and we have
\beq
\bu(\bx,t) = \bu_0[\bq(\bx,t)] = \frac{\bx-\bq(\bx,t)}{t} .
\label{vxv0q}
\eeq
On the other hand, in case there are several degenerate solutions
to (\ref{psixpsi0q})  a shock has formed at position $\bx$ and the velocity is discontinuous
(as seen from 
Eq.(\ref{vxv0q}), as we move from one solution $\bq_-$ to another one $\bq_+$ 
when we go through $\bx$ from one side of the shock surface to the other side) 
while the density is infinite.

\begin{figure}
\begin{center}
\epsfxsize=7.5 cm \epsfysize=5 cm {\epsfbox{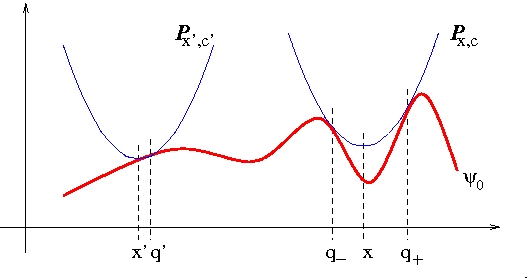}}
\end{center}
\caption{The geometrical interpretation of the Hopf-Cole solution in terms of parabolas
for the $d=1$ case.}
\label{figparabola_1d}
\end{figure}

The solution (\ref{psixpsi0q}) has a nice 
geometrical interpretation in terms of paraboloids \cite{Burgersbook,Bec2007}.
Thus, let us consider the family of upward paraboloids $\cP_{\bx,c}(\bq)$ 
centered at $\bx$ and of height $c$, with a curvature radius $t$, 
\beq
\cP_{\bx,c}(\bq)=\frac{|\bq-\bx|^2}{2t}+c .
\label{Paraboladef}
\eeq
Then, moving down $\cP_{\bx,c}(\bq)$ from $c=+\infty$, where the paraboloid is
everywhere well above the initial potential $\psi_0(\bq)$ (this is possible
for the initial conditions (\ref{ndef}) since we have $|\psi_0(\bq)| \sim
q^{(1-n)/2}$, which grows more slowly than $q^2$ at large distances),
until it touches the surface defined by $\psi_0(\bq)$, the abscissa $\bq$ of this
first-contact point is the Lagrangian coordinate $\bq(\bx,t)$. If first-contact
occurs simultaneously at several points there is a shock at the Eulerian 
location $\bx$. One can build in this manner the inverse Lagrangian map
$\bx\mapsto\bq(\bx,t)$. We show in Fig.~\ref{figparabola_1d}  an illustration
of this geometrical construction in the one-dimensional case.

\section{Lagrangian potential and matter density field}
\label{Lagrangian-potential}

\subsection{Finite viscosity $\nu>0$}
\label{Finite-viscosity}

Alternatively, the Burgers dynamics can be entirely re-expressed in terms of
Lagrangian-Eulerian mappings \cite{Vergassola1994,Bec2007}, with the introduction
of a time dependent Lagrangian potential $\varphi(\bq,t)$. 
As we shall discuss below, this also allows a precise
description of the dynamics from a Lagrangian point of view, a precious perspective
when one is interested in the evolution of the matter distribution.

Note that the inviscid dynamics (\ref{psixpsi0q}) is defined as the limit of solution
(\ref{Hopf1}) at $\nu\rightarrow 0^+$,
which is different from setting $\nu=0$ in the right hand side
of Eq.(\ref{Burgers}) (that would correspond to the so called  Zeldovich approximation
\cite{Zeldovich1970} in a cosmological context, which leads
to multi-flow regions if we first go to Lagrangian coordinates).
Thus, the inviscid limit is singular in this sense and it is useful
to start with $\nu>0$ to remove possible ambiguities that would arise by working with
Eq.(\ref{psixpsi0q}) from the onset.
Therefore, following \cite{Vergassola1994}, but for the case of nonzero $\nu$, let us
introduce the function $H(\bx,t)$, defined as
\beq
H(\bx,t) = \frac{|\bx|^2}{2} + t \psi(\bx,t) .
\label{Hdef}
\eeq
From Eq.(\ref{Hopf1}) it also reads as
\beq
H(\bx,t) = 2\nu t \ln\int\frac{\dd\bq}{(4\pi\nu t)^{d/2}} \, 
e^{[\bx\cdot\bq-|\bq|^2/2+t\psi_0(\bq)]/(2\nu t)} .
\label{Hxnu1}
\eeq
Next, as in \cite{Gurbatov1991,Saichev1996},
let us define for any function $A(\bq)$ its ``mean''
\beq
\llag A \rrag_{\bx,t;\psi_0} \equiv \frac{\int\dd\bq \, A(\bq) \,
e^{[\bx\cdot\bq-|\bq|^2/2+t\psi_0(\bq)]/(2\nu t)}}
{\int\dd\bq \, e^{[\bx\cdot\bq-|\bq|^2/2+t\psi_0(\bq)]/(2\nu t)}} ,
\label{meanA}
\eeq
seen as the average of $A$ over a probability distribution for the random variable
$\bq$, at fixed $\bx$ and $t$ for a given realization of the potential $\psi_0$.
This is possible since the weight in the average (\ref{meanA}) is positive and
normalized to unity. In particular, we have from Eq.(\ref{Hxnu1})
\beq
\llag \bq \rrag_{\bx,t;\psi_0} = \frac{\pl H}{\pl\bx} .
\label{qx_Hx}
\eeq
Note that $\llag \bq \rrag_{\bx,t;\psi_0}$ is not the Lagrangian
coordinate of the particle located at $\bx$ at time $t$.
Next, the second derivatives of $H(\bx)$ writes as
\beqa
\frac{\pl^2 H}{\pl x_i\pl x_j}  & = & \frac{1}{2\nu t} \left[ \llag q_i q_j \rrag
- \llag q_i \rrag \llag q_j \rrag \right] \nonumber \\
& = & \frac{1}{2\nu t} \llag \left( q_i - \llag q_i \rrag \right)
\left( q_j - \llag q_j \rrag \right) \rrag .
\label{HessianH}
\eeqa
Therefore, in dimension $d$, the Hessian matrix of $H(\bx)$ is the covariance matrix of
the $d$ variables $q_1,...,q_d$, which implies that it is positive semi-definite
\cite{Saichev1996}.
Moreover, it has vanishing eigenvalues if and only if the distribution of the vector $\bq$ is
degenerate (it occurs for instance when the initial velocity potential is separable,
$\psi_0(\bq)= \sum_i \psi_0^{(i)}(q_i)$, and $\psi_0^{(i)}=\psi_0^{(j)}$ for some
pair $i\neq j$). For generic potentials $\psi_0(\bq)$, such as those obtained from the
random Gaussian initial conditions (\ref{Ptheta0def}), the Hessian determinant of
$H(\bx)$ is strictly positive, which implies that $H(\bx)$ is strictly convex.

This latter property is crucial. It indeed allows to explicitly invert the function
$\bq(\bx,t)\equiv\llag \bq \rrag_{\bx,t;\psi_0}$
defined in Eq.(\ref{qx_Hx}). Thus, let us introduce the Legendre transform,
$\varphi(\bq,t)$, of $H(\bx,t)$ as
\beq
\varphi(\bq,t) \equiv \cL_{\bq} [ H(\bx,t) ] = \max_{\bx} \left[ \bq\cdot\bx 
-  H(\bx,t) \right] ,
\label{varphidef}
\eeq
where we used the standard definition of the Legendre-Fenchel conjugate $f^*(\bs)$
of a function $f(\bx)$,
\beq
f^*(\bs) \equiv \cL_{\bs} [ f(\bx) ] = \sup_{\bx} [ \bs\cdot\bx - f(\bx) ] .
\label{Legendredef}
\eeq
Because $H(\bx)$ is strictly convex, $H(\bx,t)$ is also the Legendre transform of
$\varphi(\bq,t)$ and the correspondence $\bq\leftrightarrow\bx$
associated with the Legendre transforms is one-to-one. For a fixed value of $\bq$ its
corresponding value $\bx(\bq)$ is that which makes $\bq\cdot\bx-H(\bx)$ maximal
so that $\bq$ and $\bx$ obey the relation,
\begin{equation}
\bq(\bx,t)=\frac{\partial H}{\partial \bx}.
\end{equation}
As a result $\bq(\bx)$ is given by Eq.(\ref{qx_Hx}) whereas $\bx(\bq)$ is given
by
\beq
\bx(\bq,t) = \frac{\pl\varphi}{\pl\bq} .
\label{xq_phiq}
\eeq
Finally, note that $\varphi(\bq)$ is strictly convex since the Hessian
matrix $(\pl^2\varphi/\pl q_i\pl q_j)=(\pl x_i/\pl q_j)$ is the inverse of the matrix
$(\pl q_i/\pl x_j)= (\pl^2 H/\pl x_i\pl x_j)$, which is positive definite, as seen above
from Eq.(\ref{HessianH}).

We are now in position to compare the evolution with time of the mapping $\bq\mapsto\bx$
introduced above with actual particle trajectories. First, we can note that from the
definition of the velocity potential (\ref{thetadef}) and Eq.(\ref{Hdef}) the Eulerian
velocity field also writes as
\beq
\bu(\bx,t) = - \frac{\pl\psi}{\pl\bx} = \frac{1}{t} \left[ \bx-\frac{\pl H}{\pl\bx} \right]
= \frac{\bx-\llag\bq\rrag_{\bx,t;\psi_0}}{t} ,
\label{ux_qx}
\eeq
where we used Eq.(\ref{qx_Hx}). Note the similarity with Eq.(\ref{vxv0q})
obtained for the inviscid case at regular points. However, in the viscous case
$\bu(\bx,t)$ is a priori different from the initial velocity $\bu_0[\llag\bq\rrag]$.
It is interesting to note that we also have the identity
\beq
\bx= \llag\bq\rrag + t \llag\bu_0(\bq)\rrag ,
\label{xq_u0}
\eeq
which arises from a total derivative with respect to $\bq$ of the integrand in
Eq.(\ref{meanA}), whence
\beq
\bu(\bx,t) = \llag \bu_0(\bq)\rrag_{\bx,t;\psi_0} .
\label{uxu0q}
\eeq
However, this does not imply that the Eulerian velocity field is set by the initial
velocity of Lagrangian particles that are located at $\bx$ at time $t$. In particular,
we have $\llag\bu_0(\bq)\rrag \neq \bu_0(\llag\bq\rrag)$.
Nevertheless, we can already see from Eqs.(\ref{ux_qx})-(\ref{uxu0q}) that in the
inviscid limit, $\nu\rightarrow 0^+$, and in regular regions, the ``mean'' (\ref{meanA})
becomes
dominated by a single point $\bq$ and the variance $\llag \bq^2\rrag_c$ vanishes,
in agreement with Eq.(\ref{HessianH}). We shall then recover the equation of motion of
free Lagrangian particles. Of course, this does not hold within ``shocks'', where
the ``mean'' (\ref{meanA}) is degenerate and receives contributions from several
points $\bq$.

Second, taking the derivative with respect to time of the implicit Eq.(\ref{qx_Hx}) at
fixed $\llag\bq\rrag$ yields the system of equations
\beq
1\leq i \leq d : \;\;\; 0 = \frac{\pl^2 H}{\pl x_i \pl x_j} \frac{\pl x_j}{\pl t} 
+ \frac{\pl^2 H}{\pl x_i\pl t} ,
\eeq
with summation over repeated indices (here $j$).
Using expression (\ref{Hxnu1}) to compute the second term and Eq.(\ref{HessianH})
we obtain
\beq
0 = \frac{\pl^2 H}{\pl x_i \pl x_j} \left[ \frac{\pl x_j}{\pl t}
+ \frac{\llag q_j\rrag-x_j}{t} \right] + \nu \frac{\pl^3H}{\pl x_i\pl x_j\pl x_j} ,
\eeq
and multiplying by the inverse of the Hessian matrix, $(\pl^2 H/\pl x_i \pl x_j)^{-1}$,
gives
\beq
\left. \frac{\dd x_i}{\dd t} \right|_{\llag\bq\rrag} = \frac{\pl x_i}{\pl t} =  
u_i(\bx,t) - \nu \left(\frac{\pl^2 H}{\pl x_{i'}\pl x_{j'}}\right)^{-1}_{ij} \cdot
\frac{\pl^3H}{\pl x_j\pl x_k\pl x_k} ,
\label{dxdt}
\eeq
where we used Eq.(\ref{ux_qx}). In Eq.(\ref{dxdt}) the first
term simply changes notation to rewrite $\pl\bx/\pl t$ as a total derivative at fixed
$\llag\bq\rrag$, in a form that is more familiar for a Lagrangian point of view. Thus,
Eq.(\ref{dxdt}) shows that the curves $\bx(\llag\bq\rrag,t)$, seen as a function of time
at fixed $\llag\bq\rrag$, do not follow particle trajectories since their time-derivative
differs from the Eulerian velocity field $\bu(\bx,t)$.

What is then the resulting density field?
As in \cite{Saichev1996,Saichev1997}, we can take advantage of the mapping
$\bq\leftrightarrow\bx$
defined by Eq.(\ref{qx_Hx}), that is by the Legendre conjugacy (\ref{varphidef}),
to construct the distribution of matter at any time $t$. That is, starting with a uniform
initial density $\rho_0$ at $t=0$, we can {\it define} a density field at any time $t$ by
\beq
\rho(\bx,t) \equiv \rho_0 \det\left(\frac{\pl\bq}{\pl\bx}\right) 
= \rho_0 \det\left(\frac{\pl\bx}{\pl\bq}\right)^{-1} ,
\label{rhoJacob}
\eeq
which also reads as
\beq
\rho(\bx,t) = \rho_0 \det\left(\frac{\pl^2 H}{\pl x_i\pl x_j}\right)
= \rho_0 \det\left(\frac{\pl^2 \varphi}{\pl q_i\pl q_j}\right)^{-1} .
\label{rhoHvarphi}
\eeq
In Eq.(\ref{rhoJacob}) we used the fact that the determinants are positive, as shown
by expression (\ref{rhoHvarphi}) and the convexity of $H(\bx)$ and $\varphi(\bq)$,
so that we do not need to take the absolute value to compute the Jacobians.
Of course, this model for the density field conserves the total mass.
Thus, Eq.(\ref{rhoHvarphi}), built from the Legendre conjugacy associated with
Eqs.(\ref{qx_Hx})-(\ref{xq_phiq}), is the definition of the ``geometrical model'' that we
study in this article.
We use the label ``geometrical'' to refer to the fact that it is based on Legendre transforms,
which have a geometrical interpretation in terms of convex hulls as we shall discuss below.
Moreover, it is clear from Eq.(\ref{rhoHvarphi}) that one can build in this manner the
matter distribution at any time $t$ without the need to follow the evolution of the
density field over all previous times, since the convex functions $H(\bx)$ and $\varphi(\bq)$
can be computed from the Hopf-Cole solution (\ref{Hopf1}), as in Eqs.(\ref{Hxnu1}) and
(\ref{varphidef}).

Then, from the equation of motion (\ref{dxdt}), which gives the ``trajectories'' at
constant $\llag\bq\rrag$, we obtain the evolution with time of the density
field defined by Eqs.(\ref{rhoJacob})-(\ref{rhoHvarphi}), associated with this
mapping $\bx\leftrightarrow\llag\bq\rrag$. Thus, conservation of mass yields
the standard equation
\beq
\frac{\pl\rho}{\pl t}+\nabla\cdot\left(\rho \left. \frac{\dd \bx}{\dd t} 
\right|_{\llag\bq\rrag}\right) = 0 ,
\eeq
which gives with Eq.(\ref{dxdt})
\beq
\frac{\pl\rho}{\pl t}+\nabla\cdot(\rho\bu) = 
\nu \frac{\pl}{\pl x_i} \left[ \rho \left(\Theta^{-1}\right)_{ij}
\frac{\pl\Theta_{kk}}{\pl x_j} \right], 
\label{cont1}
\eeq
where we introduced the Hessian matrix, $\Theta$, of $H(\bx)$,
\beq
(\Theta_{ij})= \left( \frac{\pl^2H}{\pl x_i\pl x_j}\right).
\label{Thetadef}
\eeq
We can note that in the one-dimensional case, $d=1$, this is identical to a simple diffusive
term $\Delta\rho$, but in higher dimensions this is generically different, see
the note~\footnote{A diffusion term
reads as 
$\nu \Delta\rho = \nu \frac{\pl}{\pl x_i} \left[ \rho \left(\Theta^{-1}\right)_{kj} 
\frac{\pl\Theta_{jk}}{\pl x_i} \right]$.
Let us briefly investigate in which cases this expression and Eq.(\ref{cont1})  coincide.
The matrix $\Theta$ being symmetric it can be diagonalized by an orthogonal matrix.
In this orthonormal basis, denoting $\lambda_k$ its eigenvalues, we obtain
$ \frac{\pl}{\pl x_i} \left[ \rho \left(\Theta^{-1}\right)_{ij}
\frac{\pl\Theta_{kk}}{\pl x_j} \right]= \rho \left( \sum_{k=1}^d
\frac{\pl\ln\lambda_k}{\pl x_i}\right) \left( \sum_{k=1}^d \frac{\lambda_k}{\lambda_i} 
\frac{\pl\ln\lambda_k}{\pl x_i}\right) + \rho \sum_{k=1}^d \frac{\pl}{\pl x_i} \left(\frac{\lambda_k}{\lambda_i}\right)
\frac{\pl\ln\lambda_k}{\pl x_i} + \rho \sum_{k=1}^d \frac{\lambda_k}{\lambda_i}
\frac{\pl^2\ln\lambda_k}{\pl x_i^2}$,
and
$\Delta\rho = \rho \left( \sum_{k=1}^d \frac{\pl\ln\lambda_k}{\pl x_i}\right)^2 
+ \rho \sum_{k=1}^d \frac{\pl^2\ln\lambda_k}{\pl x_i^2} ,$
where we sum over $i$. Therefore, both quantities are equal at points $\bx$ where
we have
$\frac{\pl\lambda_k}{\pl x_i} = 0$ and $\frac{\pl^2\lambda_k}{\pl x_i^2} = 0$ if $k\neq i$, that is,
$\frac{\pl^2\llag q_k\rrag}{\pl x_k \pl x_i} = 0$ and $\frac{\pl^3\llag q_k\rrag}{\pl x_k \pl x_i^2} = 0$ if $k\neq i$.
Of course, a simple example where these conditions are identically satisfied
is the case where the initial velocity potential is separable,
$\psi_0(\bq)=\sum_i \psi_0^{(i)}(q_i)$, and the dynamics can be factorized in terms
of $d$ one-dimensional dynamics. In generic systems they are only satisfied at
peculiar isolated points (note that to obey the last condition
it is not sufficient to have a parity symmetry $x_i\leftrightarrow-x_i$).}.

As proposed in \cite{Saichev1996}, the density field defined by the mapping
(\ref{rhoJacob}) can be interpreted by a stochastic process. Indeed, with the help of an adequate 
``mean-field approximation'' one can obtain Eq.(\ref{rhoJacob}) as the density field generated by
the motion of a passive tracer that moves along the Burgers velocity field, to which
a Brownian noise of amplitude set by $\nu$ is added (this actually yields a diffusive
term $\nu\Delta\rho$). If one does not introduce such a ``mean-field approximation'',
an analysis in terms of backward stochastic differential equations yields an inverse
problem with no explicit solution, which only simplifies to
(\ref{rhoJacob}) in the inviscid limit $\nu\rightarrow 0^+$ \cite{Ginanneschi1998}.
Thus, it might be more convenient to simply define the density field through the Jacobian
(\ref{rhoJacob}), or equivalently by the modified diffusion equation (\ref{cont1}).
This provides an explicit interpretation in terms of continuum equations of motion
as well as a simple solution at any finite $\nu$ in terms of the Legendre conjugacy
(\ref{varphidef}). As described in the next section, this ``geometrical model'' has a
well-defined inviscid limit, which allows us to recover the prescription that was used in
some previous numerical works \cite{Vergassola1994}.

\subsection{Inviscid limit $\nu\rightarrow 0$}
\label{Zero-viscosity}

We now consider the inviscid limit $\nu\rightarrow 0^+$. As we recalled in
Sect.~\ref{Hopf-Cole}, the velocity potential is given by the maximum (\ref{psixpsi0q}),
which shows degenerate maxima at shock positions.
This also determines the Eulerian velocity field $\bu(\bx,t)$.
Next, we {\it define} the density field $\rho(\bx,t)$ as the limit for $\nu\rightarrow 0^+$
of the density field (\ref{rhoHvarphi}) defined in the previous section for finite
viscosity. 

As already argued in our introduction and in \cite{Saichev1996,Saichev1997}, the main reason supporting the use of the
definition (\ref{rhoHvarphi}), that is Eq.(\ref{cont1}), is that it allows an explicit
integration of the continuity equation as the density field is given by Eq.(\ref{rhoHvarphi}) -- which remains valid
in the inviscid limit as seen below -- for any 
time $t$. In other approaches one should a priori
numerically integrate the associated continuity equation over time.
We also stress that the peculiar form of Eq.(\ref{cont1}) should not be a serious disadvantage,
since outside of shocks we recover the standard continuity equation
and ``within'' shocks the Burgers dynamics is usually seen as an effective
model. 

First, let us define the ``linear'' Lagrangian potential $\varphi_L(\bq,t)$ by
\cite{Vergassola1994}
\beq
\varphi_L(\bq,t) = \frac{|\bq|^2}{2} - t \psi_0(\bq) ,
\label{phiLdef}
\eeq
so that in the linear regime the Lagrangian map, $\bq \mapsto \bx$, associated
with particle trajectories, is given by
\beq
\bx_L(\bq,t) =  \frac{\pl\varphi_L}{\pl\bq} = \bq + t \bu_0(\bq) .
\label{xL}
\eeq
Thus we recover the linear displacement field, $\bx_L-\bq= t\bu_0(\bq)$,
which is valid before shocks appear, as seen in Eq.(\ref{vxv0q}) above.
Next, introducing the function $H(\bx,t)$, defined as in the viscous case
by Eq.(\ref{Hdef}), we obtain from Eq.(\ref{psixpsi0q}) the expression
\beq
H(\bx,t) = \max_{\bq} \left[ \bx\cdot\bq - \frac{|\bq|^2}{2} + t \psi_0(\bq) \right]
= \cL_{\bx} [ \varphi_L(\bq,t) ] ,
\label{Hxphiq}
\eeq
where we recognize the Legendre transform of $\varphi_L$.
As recalled below Eq.(\ref{psixpsi0q}) this gives the inverse Lagrangian map
$\bx \mapsto \bq$, $\bq(\bx,t)$ being the point where the maximum in
Eq.(\ref{psixpsi0q}) or (\ref{Hxphiq}) is reached. Thus, $\bq$ and $\bx$ are
Legendre-conjugate coordinates. From the previous section this is also
the inviscid limit of the mapping $\bx\leftrightarrow\llag\bq\rrag$ introduced
at finite viscosity, and outside of shocks the direct Lagrangian map, $\bq\mapsto
\bx(\bq,t)$, corresponds to particle trajectories.
This mapping is still given by  Eqs.(\ref{qx_Hx}) and (\ref{xq_phiq}), which read as
\beq
\bq(\bx,t) = \frac{\pl H}{\pl\bx} , \;\;\; \bx(\bq,t) =  \frac{\pl\varphi}{\pl\bq}  ,
\label{qx_xq}
\eeq
where the function $\varphi(\bq,t)$ is still defined by the Legendre transform
(\ref{varphidef}). From Eq.(\ref{Hxphiq}) and elementary properties of Legendre
transforms this implies that $\varphi(\bq,t)$ is also the convex hull of the
linear potential $\varphi_L(\bq)$,
\beq
\varphi(\bq,t) = \cL_{\bq} [ H(\bx,t) ] = \mbox{conv}(\varphi_L) .
\label{phi_convexhull}
\eeq
Of course, both functions $H(\bx,t)$ and $\varphi(\bq,t)$ remain convex in the
inviscid limit, in agreement with the fact that they are still given by Legendre
transforms, but they are not necessarily strictly convex (i.e. their Hessian determinant
may vanish at some points). By Legendre duality, loss of convexity in one of these
functions (i.e. hyperplanar facets) is associated with singular points (with no
well-defined gradient) in the other one.
This analysis shows how the prescription (\ref{qx_xq}), that was already used 
in some numerical computations \cite{Vergassola1994}, corresponds to the inviscid
limit of the density field (\ref{rhoJacob}) and of the mapping
(\ref{qx_Hx}), (\ref{xq_phiq}), introduced at finite $\nu$ in the previous section
(see the note \footnote{Rather than using the prescription (\ref{qx_xq}), some
numerical simulations
try to infer the distribution of matter directly from the Hopf-Cole solution. For instance,
\cite{Melott1994,Weinberg1990} first compute the velocity field over a grid of times
from the explicit solution (\ref{Hopf1}), with a small finite viscosity $\nu$, and second
integrate the particle orbits along this velocity field. 
Alternatively, working directly in the inviscid limit (\ref{psixpsi0q}),
\cite{1990MNRAS.242..200K,1992ApJ...393..437K} use the paraboloid construction
(\ref{Paraboladef}) to construct the ``skeleton'' of the Eulerian-space density field
(i.e. the Voronoi cells discussed in sect.~\ref{tessellations-2d} in this paper).
Thus, in 2D, they classify the Eulerian-space positions $\bx$ (on a grid) in three
classes, (1) regular points where there is only one first-contact point $\bq$ between
the paraboloid and the initial potential $\psi_0$, (2) singular points with two
simultaneous first-contact points, and (3) singular points with three first-contact points.
Points (2) make straight segments, connected at vertices (3), that build a Voronoi-like
tessellation with cells that contain the regular points (1). Then, from class (1) one
identifies regular regions in Lagrangian space, the complement of which defining
the singular regions which have fallen into shocks (i.e. the filaments and nodes of the
``skeleton''). Next, shocked matter is distributed on the skeleton by following the
motion in the regular regions (blue arrows in Fig.~\ref{figcoll_2d} in this paper).
This actually corresponds to an indirect implementation of the prescription
(\ref{qx_xq}) used in this article.}).

Thus, Eqs.(\ref{Hxphiq}) and (\ref{phi_convexhull}), which give the convex functions
$H(\bx,t)$ and $\varphi(\bq,t)$, and Eq.(\ref{qx_xq}) which gives the Lagrangian to
Eulerian space mapping, define the ``geometrical model'' that we study in this paper,
in the inviscid limit. This provides a simple model for the evolution with time of
the distribution of matter, which is coupled to the Burgers equation for the velocity field.
This model has already been proposed and studied in previous works 
\cite{Gurbatov1991,Saichev1996,Vergassola1994}, and the main goal of the previous
sections was to clearly set out its formulation in terms of Legendre transformations
and convex hulls, for both finite and infinitesimal viscosity. This also allowed us to
derive the modified continuity equation (\ref{cont1}) to which it corresponds.
In the following we shall investigate some of the properties of this ``geometrical model'',
focusing on the regime where all the matter is contained within shocks, which holds
as soon as $t>0$ for the power-law initial conditions (\ref{ndef}).

The correspondence $\bq\leftrightarrow\bx$ is one-to-one as long as
$\varphi_L(\bq)$ is smooth and strictly convex (which implies that $H(\bx)$
obeys the same properties). For generic initial conditions with an ultraviolet cutoff
this holds at early times, as shown by Eq.(\ref{phiLdef}).
At late times (or also at small scales, for the
power-law initial conditions that we consider in this paper), fluctuations of the initial
potential $\psi_0$ can make the linear Lagrangian potential $\varphi_L$ non-convex.
This corresponds to the formation of shocks and the maximum (\ref{Hxphiq})
is degenerate. In one dimension, $d=1$, there are generically two degenerate
Lagrangian points, $q_-<q_+$, which gives rise to a shock at Eulerian position $x_s$.
This yields a discontinuity at $x_s$ in the slope $\pl H/\pl x(x_s^{\pm})=q_{\pm}$ and
a discontinuity in the velocity $u_{\pm}=(x_s-q_{\pm})/t$.
Moreover, since particles cannot cross each other (as seen from the convexity of $H(x)$
and $\varphi(q)$, which implies that $q(x)$ and $x(q)$ are monotonically increasing),
all the particles initially located in the interval $]q_-,q_+[$ map to the shock
position $x_s$.

We can check that this gives for the Lagrangian potential $\varphi(q)$, defined by
$x(q)=\pl\varphi/\pl q$ as in the second Eq.(\ref{qx_xq}), the expression
$\varphi(q)=\varphi_L(q_-)+x_s(q-q_-)$ over $]q_-,q_+[$. Therefore,
$\varphi(q)$ is indeed the convex envelope $\varphi$ of $\varphi_L$
\cite{Vergassola1994}.
Thus, we can see that in the one-dimensional case the knowledge of the
inverse Lagrangian mapping, $x\mapsto q$, outside of shocks (whence at
their boundaries) is sufficient to reconstruct the full direct Lagrangian map
$q\mapsto x$. This also means that defining the density field as the inviscid
limit of Eq.(\ref{cont1}) (or equivalently in $d=1$ with a standard diffusive term
$\Delta\rho$), or using the standard continuity equation, give the same
results. As noticed above, this is no longer the case in higher dimensions: the inverse
map, $\bx\mapsto\bq$, is no longer sufficient to define the direct map, $\bq\mapsto\bx$,
and one must explicitly define the model used for the transportation of matter.

\subsection{``Effective momentum'' conservation}
\label{Momentum}

It is interesting to note that the ``effective momentum'' defined by the velocity field
$\bu(\bx,t)$ is conserved in a {\it global} sense in the inviscid limit by the
``geometrical model''.
Indeed, let us consider the total ``momentum'' over an Eulerian volume $\cV_{\bx}$,
\beq
\bP = \int_{\cV_{\bx}} \dd\bx \, \rho(\bx) \, \bu(\bx) .
\label{Idef}
\eeq
Here we consider a finite viscosity, $\nu>0$, so that the velocity and density fields
are regular and the integral (\ref{Idef}) is well defined.
Since $\bu$ is not the actual velocity of Lagrangian particles, as explained in
sect.~\ref{Finite-viscosity}, we refer to the quantity defined in Eq.(\ref{Idef}) as
an ``effective momentum''.
Using the expression (\ref{rhoJacob}) of the density field as the Jacobian of the
mapping $\bx\leftrightarrow\bq$, and the result (\ref{ux_qx}), we obtain
\beq
\bP = \rho_0 \int_{\cV_{\bq}} \dd\bq \, \frac{\bx(\bq)-\bq}{t}
= \frac{\rho_0}{t} \int_{\cV_{\bq}} \dd\bq \, \left( \frac{\pl\varphi}{\pl\bq} - \bq \right) ,
\label{Iq}
\eeq
where $\cV_{\bq}$ is the Lagrangian space volume which corresponds to
$\cV_{\bx}$ through the mapping $\bx\leftrightarrow\bq$.
Then, considering for instance the direction 1 in Cartesian coordinates,
we can integrate over $q_1$ as
\beq
P_1 = \frac{\rho_0}{t} \int \dd q_2 .. \dd q_d \, \left[ \varphi(q_{1+},..) -
\varphi(q_{1-},..) - \frac{q_{1+}^2 - q_{1-}^2}{2} \right] ,
\label{I1}
\eeq
where $q_{1\pm}(q_2,..,q_d)$ are the boundaries of the volume $\cV_{\bq}$
at fixed $(q_2,..,q_d)$ (and we integrate over the projection of the volume
$\cV_{\bq}$ on the $(d-1)$ remaining directions). Here for simplicity we assumed
a convex volume (so that there are only two boundaries $q_{1\pm}$) but the calculation
straightforwardly extends to the general case.
The expression (\ref{I1}) holds for any viscosity $\nu$. Then, in the inviscid limit,
$\nu\rightarrow 0^+$, the Lagrangian potential is given by the explicit expression
(\ref{phi_convexhull}) as the convex hull of $\varphi_L$.
If shocks are restricted to a finite domain $\cV_{\bx}^{\rm shock}$, which
corresponds to the finite-mass Lagrangian domain $\cV_{\bq}^{\rm shock}$,
we can evaluate the total ``effective momentum'' $\bP$ over a larger volume 
$\cV_{\bx} \supset \cV_{\bx}^{\rm shock}$, as
\beq
P_1 = - \rho_0 \int \dd q_2 .. \dd q_d \, \left[ \psi_0(q_{1+},..) -
\psi_0(q_{1-},..) \right] ,
\label{I1psi0}
\eeq
where we used Eq.(\ref{phiLdef}), since $\varphi=\varphi_L$ at the surface of
$\cV_{\bq}$. Next, using the definition of the initial velocity potential in
Eq.(\ref{thetadef}), we recognize in Eq.(\ref{I1psi0}) the $1-$component of the
initial momentum $\int \dd \bq \, \rho_0 \bu_0$.
Therefore, in the inviscid limit the ``effective momentum'' is conserved in any dimension,
even after shocks have formed, provided one considers the total momentum
over a volume such that its boundary is regular (i.e. no shock crosses its boundary).
Thus, there is no ``dissipation'' of the ``effective momentum'' in the inviscid limit.

In the general case, the ``effective momentum'' defined as the inviscid limit of the
expression (\ref{Idef}) may differ from the true momentum $\int \dd\bx \, \rho \bv$,
where $\bv$ is the actual velocity of Lagrangian particles, if shocks have formed.
Indeed, in regular regions $\bu\rightarrow\bv$ in the inviscid limit, but along shock
lines where $\bu(\bx,t)$ becomes discontinuous these velocities are not identical.
(Moreover, $\bv$ depends on whether one uses the ``geometrical model'' or a
``standard model'' based on the usual continuity equation,
as discussed in Sect.~\ref{comparison} below.)

In one dimension, where there is no ambiguity and all prescriptions for the
density field match in the inviscid limit, conservation of momentum in this limit
is a well-known property \cite{Burgersbook}.
Moreover, in this case the velocity $v$ of shock nodes is simply the mean of the
left and right velocities $u(x_-)$ and $u(x_+)$, and the ``effective momentum'' 
(\ref{Idef}) is also equal to the actual momentum of Lagrangian particles
(both are equal to $\int \dd q \, \rho_0 u_0$).

However, while in one dimension conservation of momentum also holds in a {\it local}
sense, that is, the momentum of a shock node is equal to the sum of the initial momenta
of the particles it contains, in higher dimensions this is no longer the case as shocks
can redistribute momentum among themselves. 
We shall come back to the conservation of momentum in sect.~\ref{Momentum-2d}
below, when we discuss the ``late-time'' regime where all the matter has been
redistributed over shock nodes and it is not possible to draw volumes with regular
boundaries.

\subsection{Late time structures}
\label{LateTimeEvol}

If one lets the system evolve for a long enough time, shocks generically form that lead
to the formation of low dimension structures. Eventually, finite regions $\cV_{\bq}$ in $\bq$-space can map to
a single positions $\bx_s$, leading to a point-like massive objects in the Eulerian density field. And finally, 
one expects that, at late enough time, almost all the fluid particles have reached such objects.
Conversely, it implies that the $\bx$-space is partitioned into a set of finite domains, $\cV_{\bx}$, each of them
originating from  infinitely small Lagrangian region, $\bq_v$, centered over a discrete number of points.
This gives rise to ``voids'' in Eulerian space, as the infinitesimal mass associated with
Lagrangian position $\bq_v$ is spread over the finite volume $\cV_{\bx}$
\cite{Gurbatov1991,Woyczynski2007} and this occurs when the initial conditions show significant UV power and $\psi_0(\bq)$
has a local maximum at $\bq_v$ with a discontinuous first-derivative.
In such regions, the function $H(\bx,t)$ is affine, with a constant gradient $\pl H/\pl\bx=\bq_v$,
over the domain $\cV_{\bx}$.

This is the regime we investigate here. 
More precisely, we assume that $\varphi_{L}(\bq)$ shows infinitesimally thin downward
peaks over a set of peak locations $\vq_{i}$, so that its convex hull is supported by a subset
of those points. It is only in the close vicinity of those points that the mapping is regular, 
and thus that the local density is there independent of the chosen model for the continuity equation.
However, almost all the mass is then in the shock manifold(s) and the question we want to address is how the mass
is distributed within those manifolds. For the 1D case the question can be easily answered as 
shocks cover disconnected point-like regions, each of them being associated with a single 
mass halo. For higher dimension cases however, the shocks form
a single connected manifold where objects of different dimension and at different locations co-exist. It appears that
the way the mass is distributed within this manifold inherently depends on which prescription is adopted for the inviscid limit 
of the continuity  equation.

Let us then explore in more details the case of the geometrical model in such a regime.
Basic properties of its Legendre transform can easily be derived as it then reads (for a fixed time $t$),
\begin{equation}
H(\vx)=\max_{\vq}[\vx\cdot\vq-\varphi_{L}(\vq)]=\max_{i}[\vx\cdot\vq_{i}-\varphi_{L}(\vq_{i})] .
\end{equation}
It is clear from this form that $H({\bf x})$ is affine over regions, corresponding to a
given $i$. Let us label them by 
the corresponding parameter $i$. Boundaries between 
two such regions, say $i$ and $j$, correspond to locations where 
\begin{equation}
\vx\cdot\vq_{i}-\varphi_{L}(\vq_{i})=\vx\cdot\vq_{j}-\varphi_{L}(\vq_{j}).
\end{equation}
In 2D this obviously corresponds to a straight line (and more generally to hyperplanes).
The $\vx$-space is then partitioned into polygons, 
in which $H({\bf x})$ is affine. This defines a tessellation, which a priori resembles that
of Voronoi (we shall see later how the two are linked).

These regions are simply ``voids'' where the matter that was in the infinitely small
$\vq$-region $i$ has 
spread. Note that $H(\vx)$ is continuous, but has discontinuous derivatives on the
boundaries of those regions, which thus correspond to shocks. Depending on the space
dimension there are shocks of various dimensions. For instance in 2D,
segments bounding two adjacent domains are shock lines. They still gather an infinitesimally
small amount of matter, 
which corresponds to that contained in the ridge joining two neighbored $i$ regions in
$\vq$-space (see \cite{Vergassola1994,Bec2007} 
for a depiction of those objects). Most of the mass has actually reached the point-like shocks
(all of the mass at any time $t>0$ in the regime that we consider in this paper).
In Lagrangian space they correspond to regions where the convex hull $\varphi(\bq)$
is of a constant slope. 

An important geometrical result is that these regions are triangles and that the triangulation
is the one associated with the tessellation we have just defined. Indeed each point-like shock
is associated with each summit $\vx_{c}$ of the tessellation. Those points are at the
intersections of 3 domains (in 2D) and there are thus three values, $i_{1}$, $i_{2}$ and
$i_{3}$, such that $\vx_{c}\cdot\vq_{i}-\varphi_{L}(\vq_{i})$ is constant. It can easily be
checked that no points inside the triangle $(\vq_1,\vq_2,\vq_3)$ 
can be part of the convex hull and the whole
region within the triangle has reached the point $\vx_{c}$. In terms of mass distribution this
means that the mass
function is given by the area distribution of triangles of that triangulation. This property
extends to higher dimensions. In 3D, the masses correspond to the volume of tetrahedra
\cite{Gurbatov1991}.

This construction shares some similarities with the classical Voronoi tessellation/Delaunay 
triangulation dual construction. Let us remind the reader that a Voronoi tessellation is built
from a set of seeds. It is the surface of a volume partition whose domains are such that all 
points of each domain are closest to a given seed. The tessellation we have in the present
``geometrical model'' is not a true Voronoi tessellation. Indeed, the $d-1$-hypersurface
defining the boundary between two domains $i$ and $j$, $\mS^{(d-1)}_{ij}$, is equivalently
defined by
\begin{equation}
\vert\vq_{i}-\vx\vert^2-2t\psi_{0}(\vq_{i})=\vert\vq_{j}-\vx\vert^2-2t\psi_{0}(\vq_{j}) .
\label{VorBoundaries}
\end{equation}
It corresponds to the hypersurface of the equidistant points of $\vq_{i}$ and $\vq_{j}$ only
when $\psi_{0}(\vq_{i})=\psi_{0}(\vq_{j})$. 
It is however possible to embed our $d$-space into a $d+1$-space such that the extra
coordinate $q^{d+1}$ of $\vq$ is $\sqrt{2t(C_{0}-\psi_{0}(\vq))}$ (where $C_{0}$ is a large
enough constant so that all these quantities are real, if we consider a finite volume over
$\bq$). Then the $d$-dimension hypersurface
of the equidistant points between $\vq_{i}$ and $\vq_{j}$ (in the $d+1$ space) is defined as
\begin{eqnarray}
\vert\vq_{i}-\vx\vert^2+\left(x^{d+1}-\sqrt{2t(C_{0}-\psi_{0}(\vq_{i}))}\right)^2&=&\nonumber\\
&&\hspace{-6.3cm}\vert\vq_{j}-\vx\vert^2+\left(x^{d+1}-\sqrt{2t(C_{0}-\psi_{0}(\vq_{j}))}\right)^2.
\end{eqnarray}
Its intersection with  the $x^{d+1}=0$ hypersurface is nothing but $\mS^{(d-1)}_{ij}$. 
It straightforwardly follows that the tessellation we have constructed is nothing but a plane
cut through a higher dimensional Voronoi tessellation. Note that  some properties of the
Voronoi tessellation have then been lost. In particular, seed points may have no domain at all
associated with (this is when they are not part of the convex hull) and the seed point of a
given domain does not necessarily sit within it.

It is interesting to note that standard Voronoi tessellations have also been used in cosmology
to study the large-scale structures of the Universe. They provide a model of these large-scale
structures which can reproduce some properties of the observed galaxy distribution
\cite{Icke1987,Weygaert1989,Weygaert2007}.
On the other hand, they can be used as data analysis tools, to measure for instance the
velocity field statistics \cite{Bernardeau1996}.
See for instance \cite{Weygaert2002,Weygaert2009} for reviews.
The facts that the Burgers dynamics leads to generalized Voronoi cells as described above,
see also \cite{Woyczynski2007}, and that this model provides a good description
of gravitational clustering at large scales in cosmology
\cite{Gurbatov1989,Weinberg1990,1992ApJ...393..437K,Vergassola1994}, provide
a further motivation for the use of Voronoi tessellations in this context.
Note that the Burgers dynamics (more precisely its formulation as the ``geometrical model''
studied here) also provides the dynamical evolution of these
tessellations, as a function of the initial conditions. In particular, this makes explicit the
connection with the gravitational dynamics and the background cosmology.

The picture we have obtained so far is a simple consequence of the convex hull construction
from which the geometrical model derives
and describes a snapshot of the mass distribution. The question we explore hereafter is how this
structure evolves with time.

\section{One-dimensional dynamics}
\label{One-dimension}

As time grows one expects the fluid particles to aggregate in halos of increasing mass.
The net result of the time evolution should then be that of merging effects. For the 1D case,
this happens simply by the merging of adjacent halos. Let us illustrate the mechanism at
play more explicitly.

\subsection{Inverse and direct Lagrangian maps}
\label{Lagrangian-maps-1d}

\begin{figure}
\begin{center}
\epsfxsize=7.5 cm \epsfysize=5 cm {\epsfbox{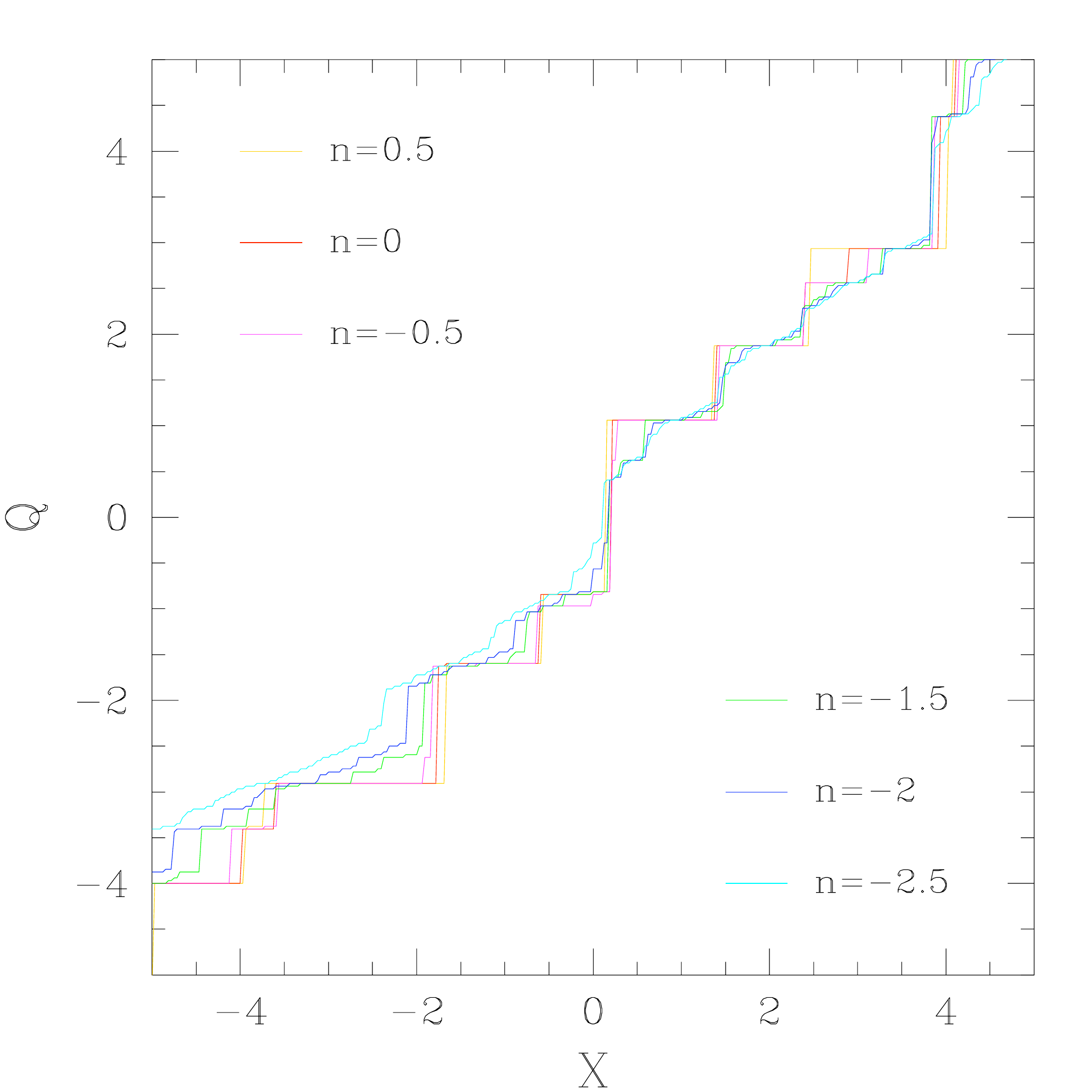}}
\end{center}
\caption{(Color online)
The inverse Lagrangian map, $X\mapsto Q$, for several initial power-spectrum index
$n$. We display the results in terms of the dimensionless scaling variables $Q$ and
$X$, so that the properties of the curve $Q(X)$ (its increments if $n\leq -1$) do
not depend on time.}
\label{figqx_1d}
\end{figure}

From the Hopf-Cole solution (\ref{psixpsi0q}), or the Legendre transform (\ref{Hxphiq}),
we obtain the Eulerian velocity field $u(x,t)$ and its potential $\psi(x,t)$,
as well as the inverse Lagrangian map, $x\mapsto q$.
As noticed above, from standard properties of Legendre transforms, and as can be
directly checked on Eq.(\ref{Hxphiq}), the map $q(x)$ is monotonically increasing.
This expresses the fact that particles cannot cross, because of the infinitesimal
viscosity $\nu$.
We plot for illustration in Fig.~\ref{figqx_1d} the results obtained for $q(x)$ for one
realization of the initial conditions (\ref{ndef}), in the six cases $n=0.5,0,-0.5,-1.5,-2$
and $-2.5$, at some time $t$.
Note that we actually display the dimensionless scaling variables $X\mapsto Q$,
defined in (\ref{QXU}), so that the time of the output is irrelevant since
the statistical properties of the curve $Q(X)$ (its increments if $n\leq -1$)
do not depend on time.
Moreover, we can check that the typical scale (e.g., size of vertical jumps or
horizontal plateaus) is indeed of order unity.

We can check that $Q(X)$ is monotonically increasing.
In agreement with previous works
\cite{She1992,Vergassola1994,AvellanedaE1995,Avellaneda1995,
Frachebourg2000,Valageas2009c},
we can see that for $-1<n<1$, which we shall
label as the ``UV class'' of initial conditions since $P_{\theta_0}(k)$ shows
significant power at high $k$, there are large voids (finite intervals over $X$ 
where $Q$ is constant) and a finite number of shocks per unit length, where
$Q(X)$ is discontinuous and shows positive jumps of order unity.
For $-3<n<-1$, which we shall label as the ``IR class'' of initial conditions since
$P_{\theta_0}(k)$ shows significant power at low $k$, we can distinguish a
proliferation of small jumps \cite{She1992,Vergassola1994}.
Indeed, for $n=-2$ it can be shown that shocks are dense
in Eulerian space so that there is an infinite number of shocks per unit length (the
shock mass function diverges at low mass)
\cite{Sinai1992,Aurell1997,Bertoin1998,Valageas2009a}.
In addition, there are still large shocks, of mass of order unity, associated with
vertical jumps of order unity.

The initial conditions used to generate Fig.~\ref{figqx_1d} have the same Fourier
phase for each velocity mode $\ut_0(k)$ for all $n$, that is, going from the case
$n_1$ to $n_2$ one only multiplies each $\ut_0(k)$ by $k^{(n_2-n_1)/2}$ to satisfy 
the power spectrum (\ref{ndef}). Moreover, the times are chosen so that a given $X$
corresponds to the same $x$ (i.e. the figure is obtained from different times with $n$,
such that they all have the same scale $L(t)$).
As noticed in \cite{Vergassola1994}, this makes the structures seen in the various cases 
very similar, roughly located at the same places. This is clearly apparent in
Fig.~\ref{figqx_1d} as we can see that the various curves $Q(X)$ roughly superpose
on each other. (For the IR class this is not always the case since structures can
move over a large distance because of the large power at long wavelengths
of the velocity field.)

\begin{figure}
\begin{center}
\epsfxsize=7.5 cm \epsfysize=5 cm {\epsfbox{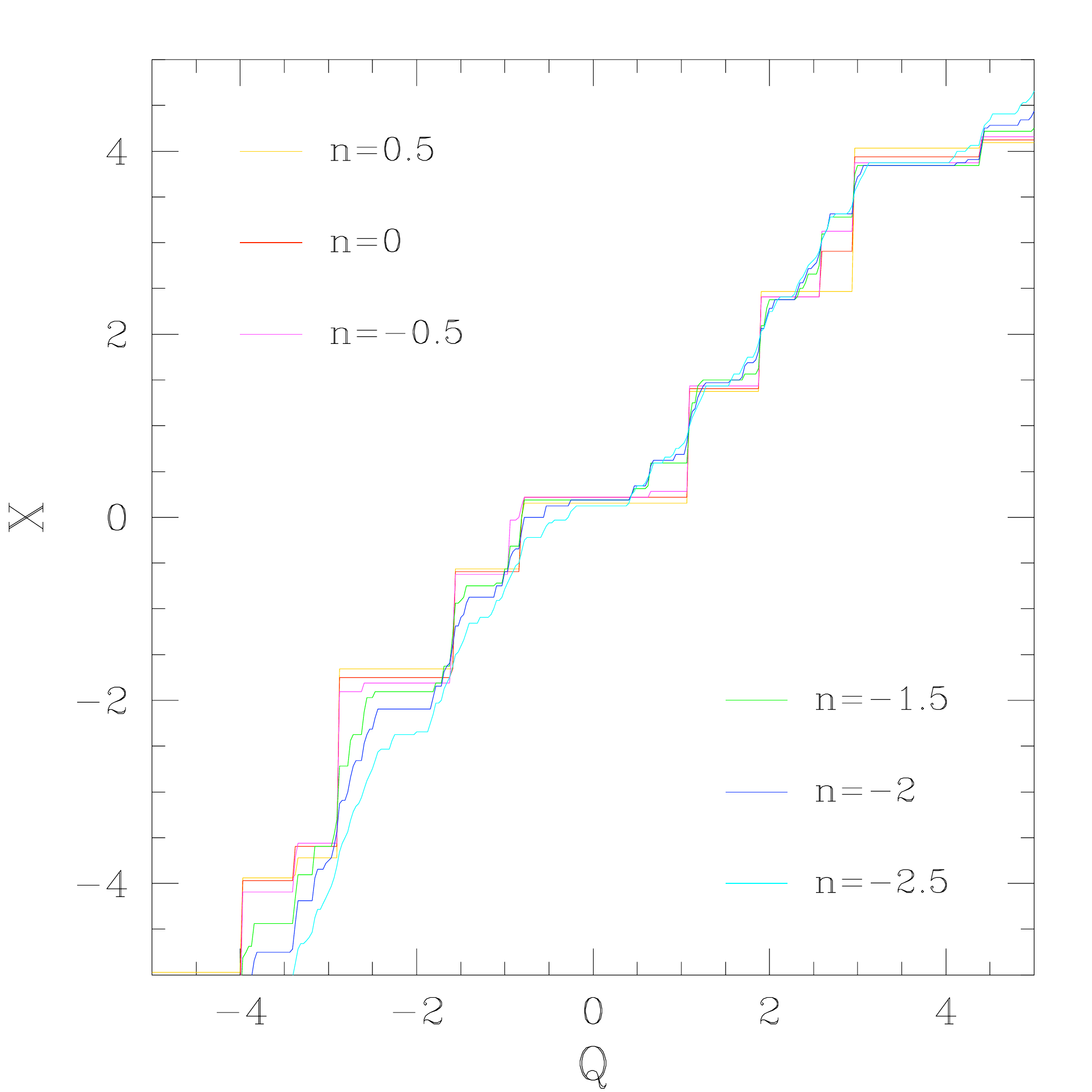}}
\end{center}
\caption{(Color online)
The direct Lagrangian map, $Q\mapsto X$, for several initial power-spectrum index
$n$.}
\label{figxq_1d}
\end{figure}

Next, to derive the direct Lagrangian map, $q\mapsto x$, we do not need to use
the second Legendre transform (\ref{phi_convexhull}) with Eq.(\ref{qx_xq}).
Indeed, as noticed in Sect.~\ref{Zero-viscosity},
in one dimension the map $q(x)$ is sufficient to reconstruct $x(q)$ since
particles do not cross each other, so that both functions $x(q)$ and $q(x)$ are
monotonically increasing \cite{Vergassola1994}.
Therefore, spanning $q(x)$ we obtain $x(q)$ (up to the resolution of our grid).
In other words, rotating Fig.~\ref{figqx_1d} by 90 degrees we can directly read $X(Q)$.
We display our results in Fig.~\ref{figxq_1d}, for the same initial conditions as
in Fig.~\ref{figqx_1d}. We clearly see the horizontal plateaus associated with shocks,
that form a Devil's staircase for $-3<n<-1$ where there is a proliferation of small
shocks, in agreement with Refs.~\cite{Vergassola1994,Aurell1997}.

\begin{figure}
\begin{center}
\epsfxsize=7.5 cm \epsfysize=5 cm {\epsfbox{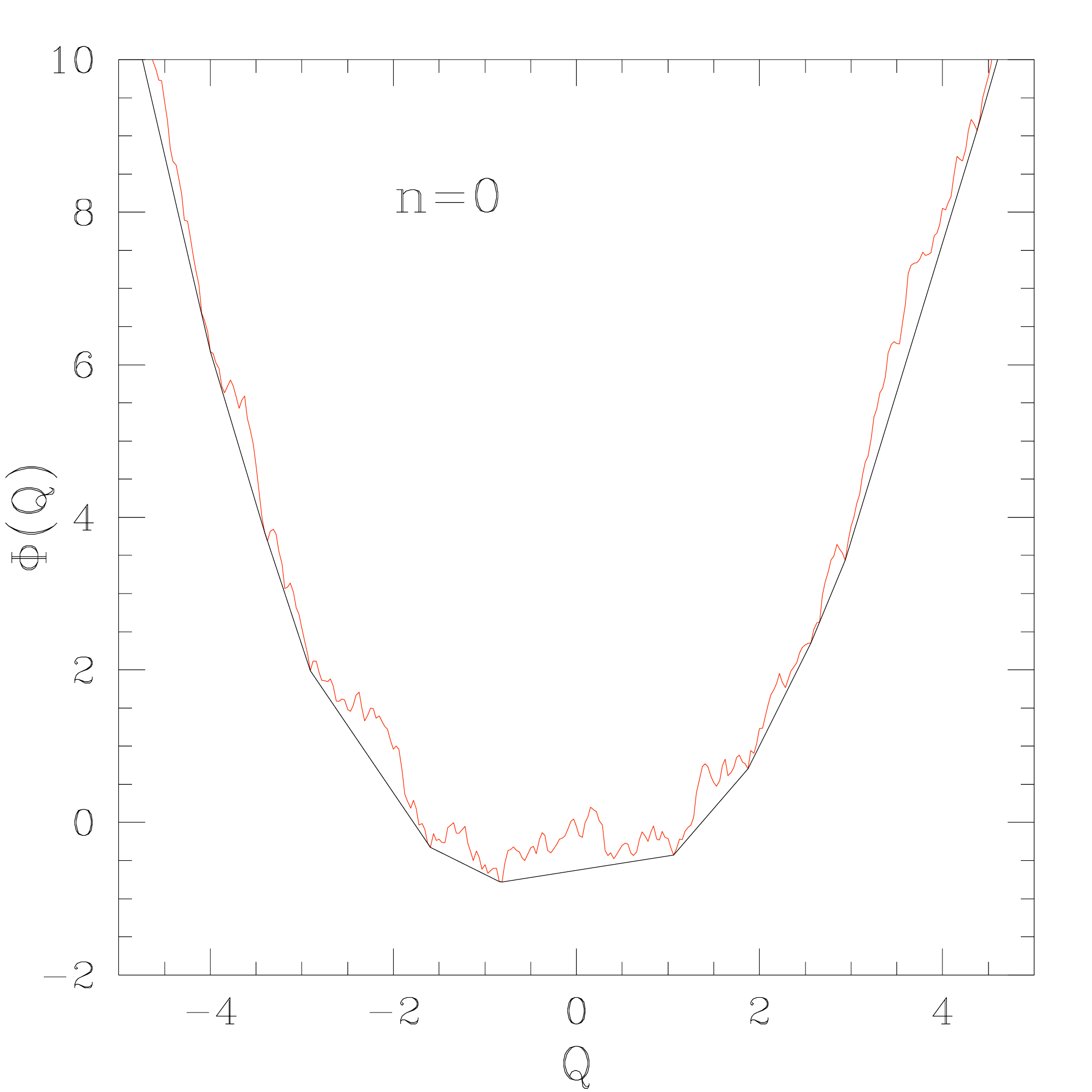}}
\epsfxsize=7.5 cm \epsfysize=5 cm {\epsfbox{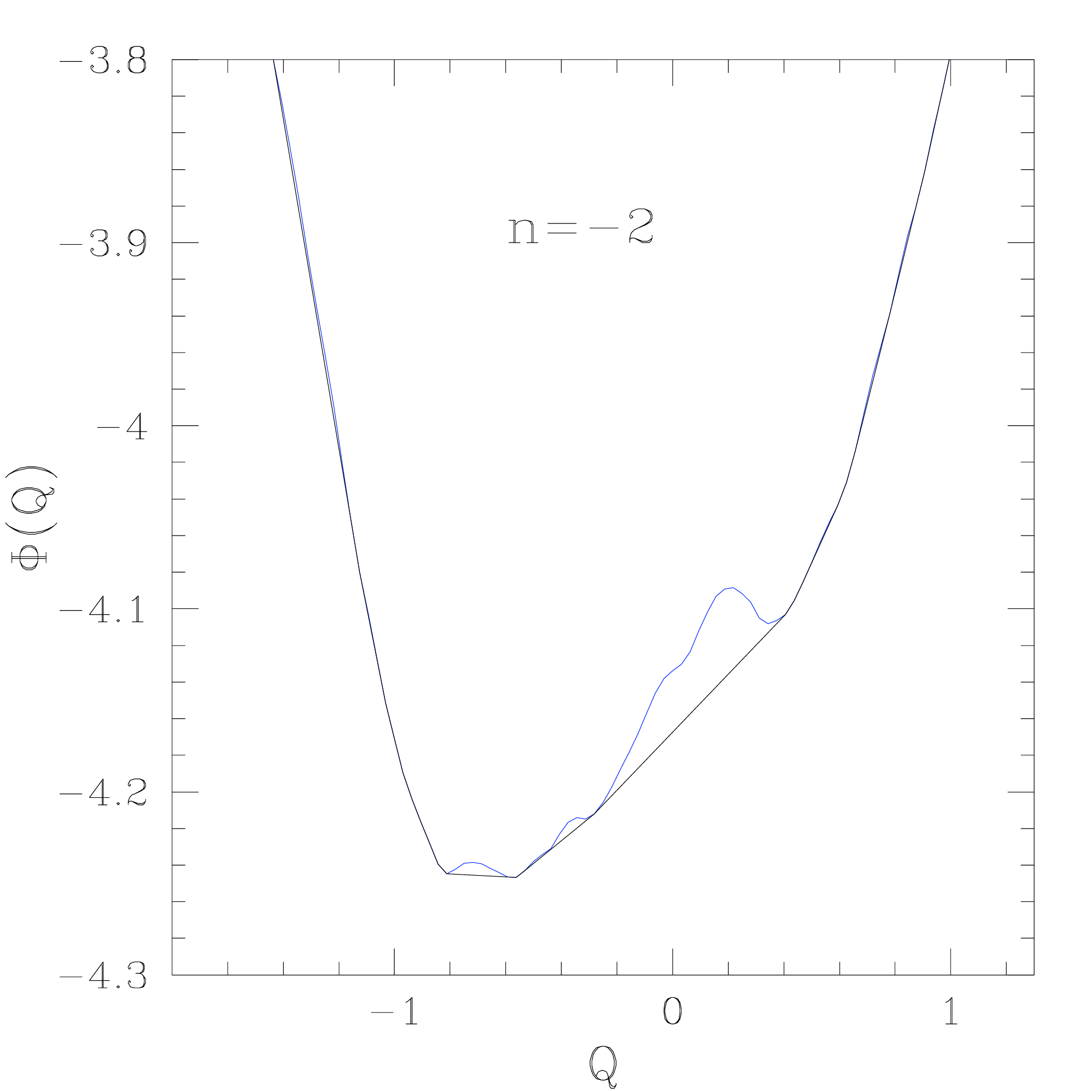}}
\end{center}
\caption{(Color online)
The linear Lagrangian potential $\varphi_L(q)$ and its convex hull $\varphi(q)$,
in terms of the dimensionless scaling variables $Q$ and $\Phi=\varphi/L(t)^2$.
We show the cases $n=0$ (upper panel) and $n=-2$ (lower panel).}
\label{figPhiq_1d}
\end{figure}

For illustration purposes, even though it is not needed to derive the Lagrangian map
$x(q)$, we display in Fig.~\ref{figPhiq_1d} the linear Lagrangian potential
$\varphi_L(q)$ and its convex hull $\varphi(q)$, for the cases $n=0$ and $n-2$, see
also \cite{Vergassola1994}.
We use the dimensionless scaling variables $Q$ and $\Phi=\varphi/L(t)^2$.
For the white-noise case, $n=0$, where $\varphi_L(q)$ has no finite first derivative,
the convex hull only touches $\varphi_L(q)$ at isolated points, which correspond
to the boundaries of shocks in Lagrangian space (and the position $x_s$ of the
shock in Eulerian space is given by the constant slope of $\varphi$ in-between
these boundaries $\{q_-,q_+\}$). Again, we can check that typical scales are of order
unity.
For the Brownian case, $n=-2$, where $\varphi_L(q)$ has no finite second derivative,
we can see some regions of Lagrangian space where the convex hull is significantly
below $\varphi_L$ over a large interval, associated with a massive shock, but
in many places it is hard to distinguish both curves. Indeed, since shocks are dense
there are finite-size regions which contain an infinite number of contact points
(note also the different scales between the upper and lower panels).

\subsection{Mergings}
\label{Mergings}

As time grows, shocks merge to form increasingly massive point-like objects,
so that their typical mass scales as $\rho_0 L(t)$. Moreover, once two particles
have coalesced into a single shock they remain glued together ever after.
Again, this can be directly seen from the Hopf-Cole solution (\ref{psixpsi0q}).
Let us have a closer look on Fig.~\ref{figparabola_1d} which illustrates the parabolic
construction
(\ref{Paraboladef}). On the left side we show the case of a regular point $x'$, associated with
a unique Lagrangian coordinate $q'$, whereas on the right side the first-contact
parabola $\cP_{x,c}$ simultaneously touches the potential $\psi_0(q)$ at the
two first-contact points $q_-$ and $q_+$. This implies the points within $]q_-,q_+[$ cannot
come into contact with a parabola. There is actually a shock at
the Eulerian position $x$, which gathers all the mass associated with the
Lagrangian interval  $]q_-,q_+[$. Then, as the parabola curvature decreases with time as
$1/t$ the points within  the interval $]q_-,q_+[$ will remain unreachable.
This implies that this interval cannot be broken into separate pieces and the 
whole interval $]q_-,q_+[$ can only belong to a single shock, which can eventually
grow larger.

\begin{figure}
\begin{center}
\epsfxsize=7.5 cm \epsfysize=5 cm {\epsfbox{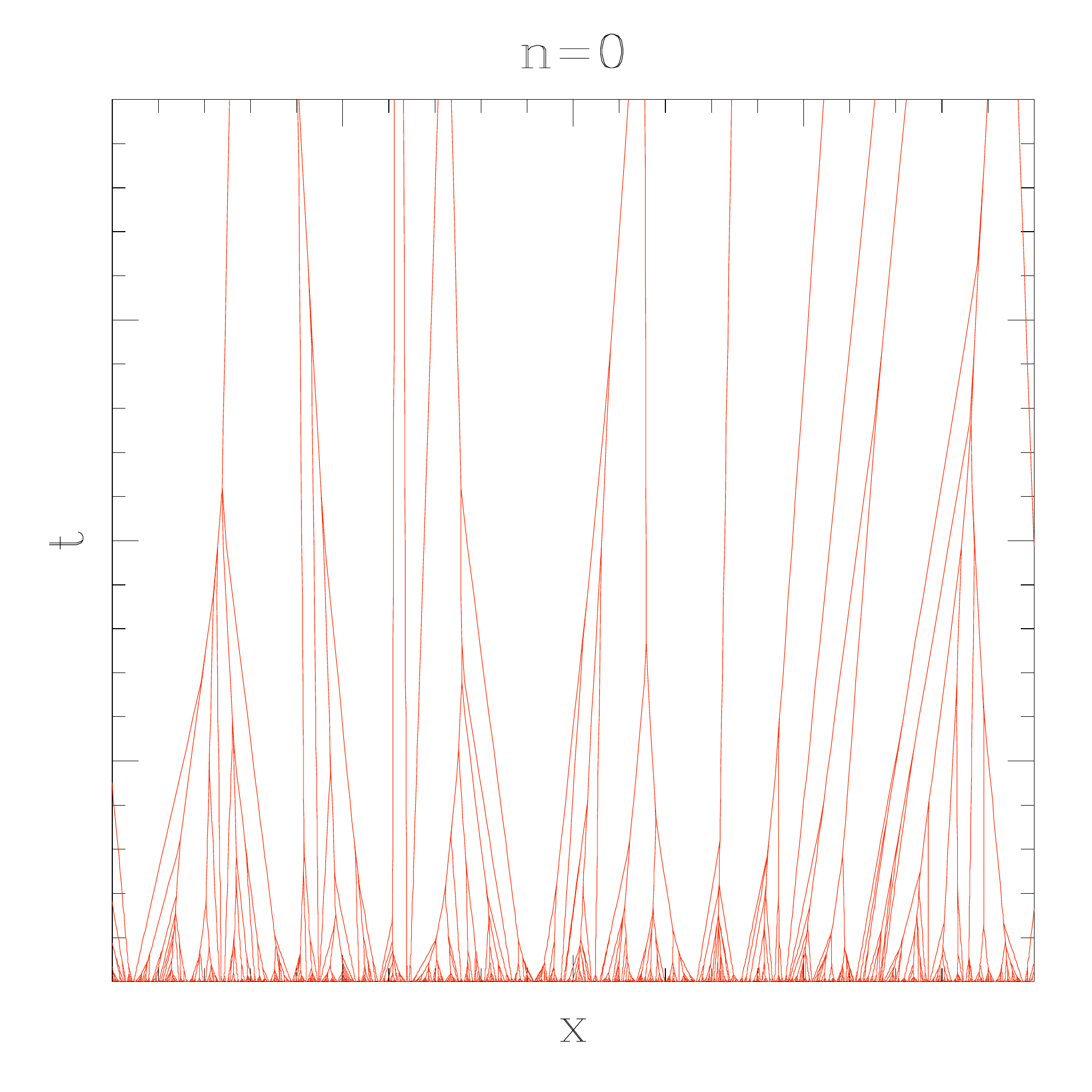}}
\end{center}
\caption{(Color online)
The particle trajectories obtained for one realization in the white-noise case $n=0$.
For any time, each line correspond to a collection of particles that have reached the same
location.}
\label{figtree_1d}
\end{figure}

We show in Fig.~\ref{figtree_1d} the particle trajectories obtained for one realization
in the case $n=0$, using the physical coordinates $x$ and $t$, see also
\cite{Noullez2002}. We can see how particles are gathered into isolated shocks that merge as
time grows to build increasingly massive and rare point-like masses. This merging process
happens when two lines meet. At such points two halos merge into a single halo. The mass
gathered in a single halo then corresponds to the volume span by all its progenitors. The
average mass obviously grows with time (according to the scaling law (\ref{selfsimilar})
for the initial conditions (\ref{ndef})).
Note that in between mergings halos move along straight lines, in agreement with 
Eq.(\ref{Burgers}). At merging point their velocity changes. Its new value corresponds to the
average of the mass weighted velocities of the halos that merge together \cite{Burgersbook}.
Note finally that in between lines there is no matter left. These are voids.

Other spectrum indices in the range $-1<n<1$ yield similar figures (with shocks getting
more numerous for lower $n$), while for $-3<n<-1$ the particle trajectories fill the whole
$(x,t)$ plane since shocks are dense in the Eulerian space. Note however that their boundary
lines are not dense in Lagrangian space. This is simply due to the fact that although their
number density is infinite, halos have finite masses.

\section{Two-dimensional dynamics}
\label{Two-dimensions}

We now consider the two-dimensional case, $d=2$. The Hopf-Cole solution and its associated
parabola construction can be transposed from the 1D case. The merging properties of the halos are
however much more intricate.

\subsection{Lagrangian and Eulerian tessellations}
\label{tessellations-2d}

In high dimensions the inverse Lagrangian map, $\bx\mapsto\bq$, is still given by
the Hopf-Cole solution (\ref{psixpsi0q}), that is, by the Legendre transform
(\ref{Hxphiq})-(\ref{qx_xq}). In the one-dimensional case, when shocks have formed, that is 
when $x$ maps to the degenerate pair $\{q_-,q_+\}$, monotonicity arguments ensure that
conversely the whole interval $]q_-,q_+[$ maps to position $x$.
For higher dimensional cases this is no more the case : the image of the function $\bq(\bx)$
does not span the whole $\bq$-space
so that the map $\bx\mapsto\bq$ does not fully determine the inverse function
$\bx(\bq)$ by itself (and a further prescription must be added).
That would be the case if the first-contact paraboloid $\cP_{\bx,c}$,
introduced in Eq.(\ref{Paraboladef}), associated with a massive shock node
at $\bx$, would make contact with the potential $\psi_0(\bq)$ over a closed curve
$\cC_{\bx}$. But this is generically not so: $\cP_{\bx,c}$ only touches the potential
$\psi_0(\bq)$ 
over a few points -- from one for a regular point to $d+1$ for a shock node. 
As described in Sect.~\ref{Lagrangian-potential}, this missing information
is provided by the Lagrangian potential $\varphi(\bq)$, through Eqs.(\ref{qx_xq})
and (\ref{phi_convexhull}), which take into account that the maps $\bq(\bx)$ and
$\bx(\bq)$ arise from the Burgers dynamics, to which we have coupled the evolution
of the matter distribution defined by the ``geometrical model''.

\begin{figure}
\begin{center}
\epsfxsize=6 cm \epsfysize=6 cm {\epsfbox{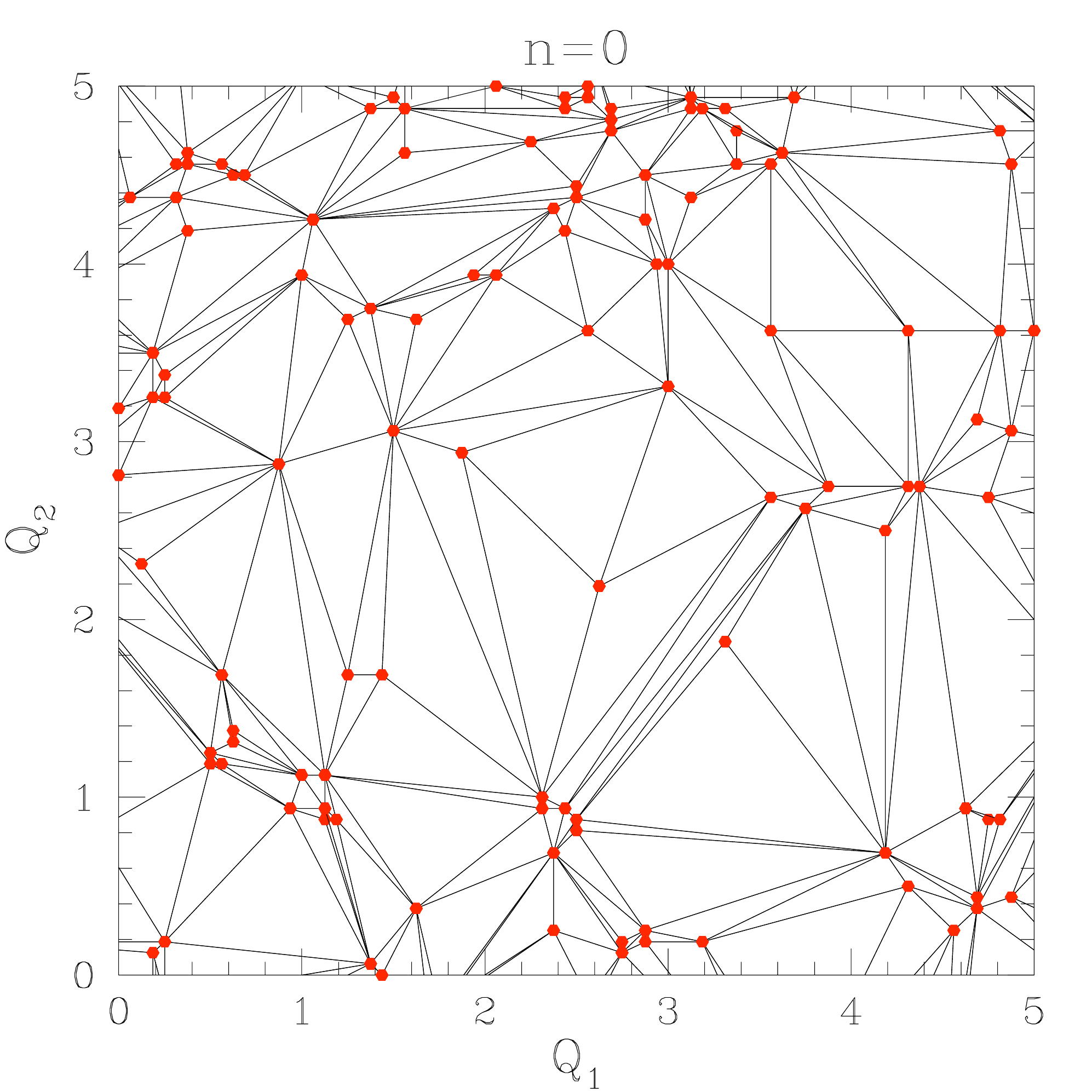}}
\epsfxsize=6 cm \epsfysize=6 cm {\epsfbox{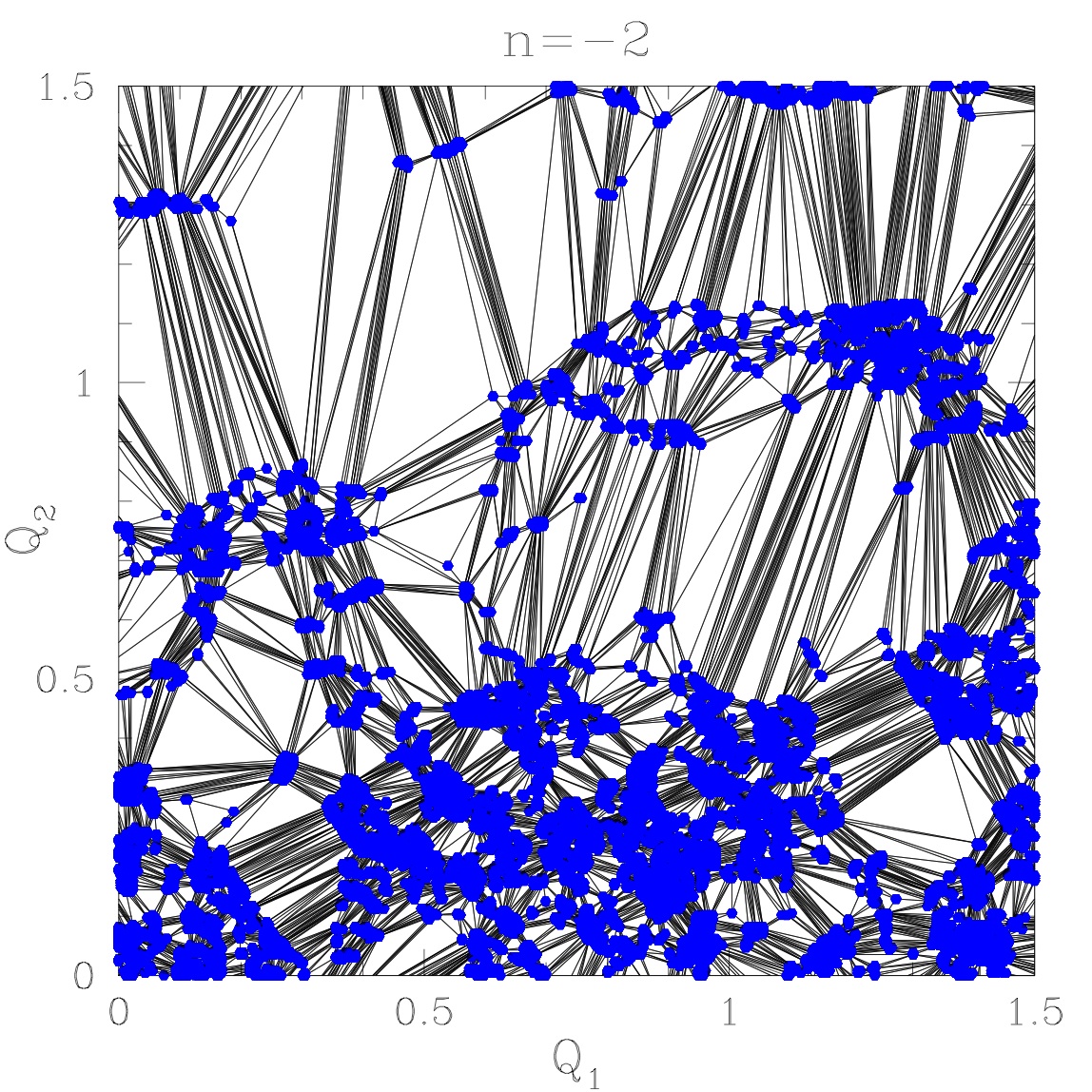}}
\end{center}
\caption{(Color online)
The partition of the Lagrangian space associated with triangular facets of the
convex hull $\varphi(\bq)$ of the linear Lagrangian potential. We plot the triangulation
in terms of the dimensionless scaling coordinates $(Q_1,Q_2)$.
The Lagrangian potential realization corresponds 
to the case $n=0$ (upper panel) and $n=-2$ (lower panel).}
\label{figtriang_2d}
\end{figure}

Since we focus in this section on the two-dimensional case,
for generic initial velocity potential $\psi_0$ the convex hull $\varphi$ 
is made of regular parts, ruled surfaces and planar triangles, see Fig.5 of
\cite{Vergassola1994}. However, for the power-law initial conditions (\ref{ndef}),
where the initial velocity potential obeys the scaling law
\beq
\lambda > 0 : \;\; \psi_0(\lambda\bq) \law \lambda^{(1-n)/2} \psi_0(\bq) ,
\label{scaling_psi0}
\eeq
so that $\varphi_L(\bq)$ shows no finite first derivative if $-1<n<1$, and no
finite second derivative if $-3<n<-1$, there are no regular parts and the convex
hull is entirely made of planar triangles. This corresponds to the late time limit described in 
Sect.~\ref{LateTimeEvol}, a regime in which all the mass is located within
shock nodes, see also \cite{Gurbatov1991}.
Note that this holds for any dimension, for the initial conditions
(\ref{ndef}), the convex hull $\varphi(\bq)$ being made of pieces of hyperplanes of
dimension $d$, defined by $d+1$ points.
This is illustrated on Fig.~\ref{figPhiq_1d} for the one-dimensional case, on 
Fig. \ref{figtriang_2d} for the two-dimensional cases.

We show in Fig.~\ref{figtriang_2d} the triangulations we obtain in the $\bq$-space
at a given time and for one realization of the initial potential $\psi_0$, for the
two cases $n=0$ (upper panel) and $n=-2$ (lower panel). We can see that we recover
the qualitative properties found in the one-dimensional case. 
In the case $n=0$ contact points of the
convex hull (which are the nodes of the triangulations) are well separated
and most triangles have an area of order unity, which implies that in Eulerian
space shock nodes are in finite number per unit surface with masses of order
unity (in the dimensionless scaling units (\ref{QXU})).
In the case $n=-2$ there are regions of the Lagrangian space with an infinite
number of nodes, with numerous triangles of very small area (up to the
numerical resolution), which implies that in Eulerian space there is also
an infinite number of shocks per unit surface (in the mean), that is, the mass function
of shocks diverges at low masses. We can also distinguish some ``white'' areas in the
figure, associated with triangles of area of order unity, that is, shock nodes with a mass
of order unity.

The Lagrangian space triangulation shown in Fig.~\ref{figtriang_2d}, which arises
from the triangular facets of the convex hull $\varphi(\bq)$, is associated
with a dual partition of the Eulerian space, which arises from the planar facets
of the Legendre transform $H(\bx)$. Indeed, as explained in
Sect.~\ref{Lagrangian-potential}, the convex functions $H(\bx)$ and $\varphi(\bq)$
are Legendre transforms of each other, and the planar facets of $\varphi(\bq)$
are associated with vertices of $H(\bx)$, given by the second Eq.(\ref{qx_xq}), whereas
vertices of $\varphi(\bq)$ (i.e. the nodes of the triangulation of Fig.~\ref{figtriang_2d},
where $\varphi_L$ makes contact with its convex hull) are associated with
planar facets of $H(\bx)$, with a slope given by the first Eq.(\ref{qx_xq}).

\begin{figure}
\begin{center}
\epsfxsize=6 cm \epsfysize=6 cm {\epsfbox{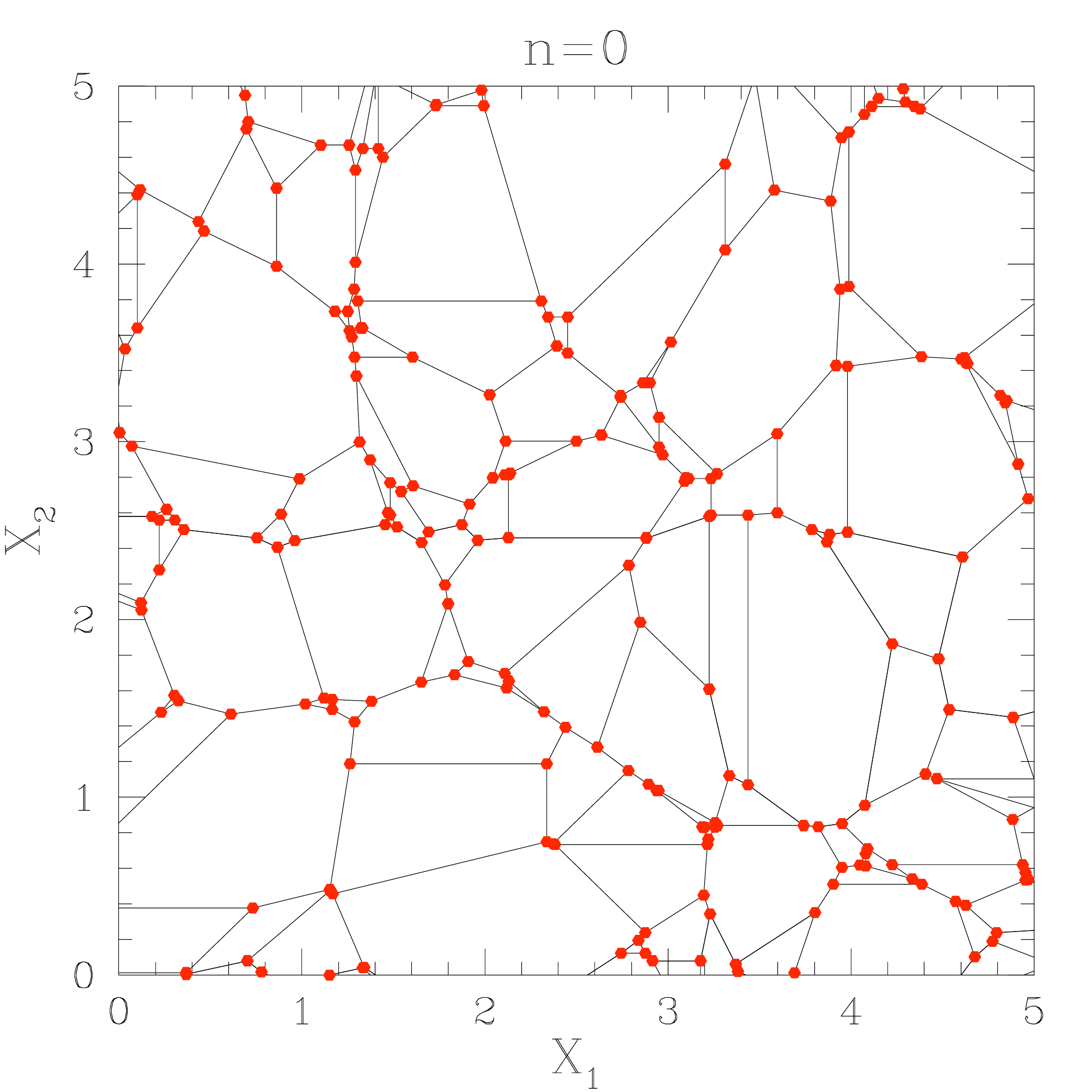}}
\end{center}
\caption{(Color online)
The ``Voronoi-like'' tessellation of the Eulerian space associated with planar facets of
the convex function $H(\bx)$. It is the dual of the Lagrangian space triangulation of
Fig.~\ref{figtriang_2d}. We plot the diagrams in terms of the dimensionless scaling
coordinates $(X_1,X_2)$ for the case $n=0$.}
\label{figvor_2d}
\end{figure}

We show in Fig.~\ref{figvor_2d} the resulting ``Voronoi-like'' tessellation of the Eulerian space
obtained for a realization of the initial potential $\psi_0$ at a given time, for the $n=0$ case. 
In agreement with the Lagrangian triangulations obtained in Fig.~\ref{figtriang_2d},
we can see that in the case $n=0$ shock nodes are isolated and in finite number per unit
area, while void sizes are of order unity.
In the case $n=-2$ cells have such a small area (which keeps decreasing as we increase the
numerical resolution) that vertices appear to cover the whole
plane, suggesting that shock nodes are dense, as in the one-dimensional case.

\begin{figure}
\begin{center}
\epsfxsize=7.5 cm \epsfysize=5 cm {\epsfbox{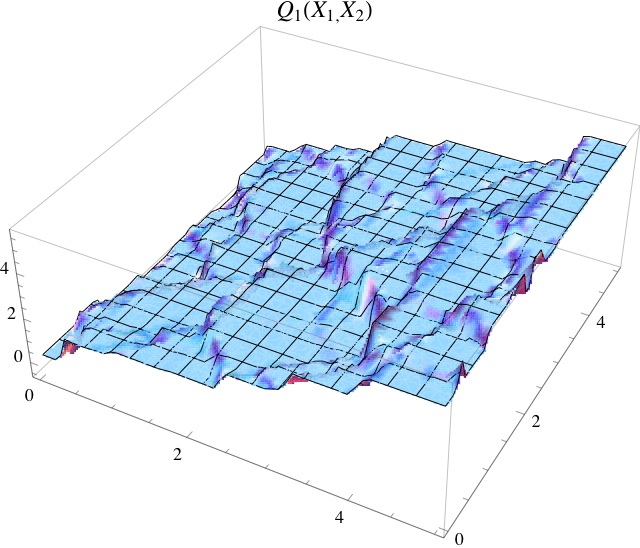}}
\epsfxsize=7.5 cm \epsfysize=5 cm {\epsfbox{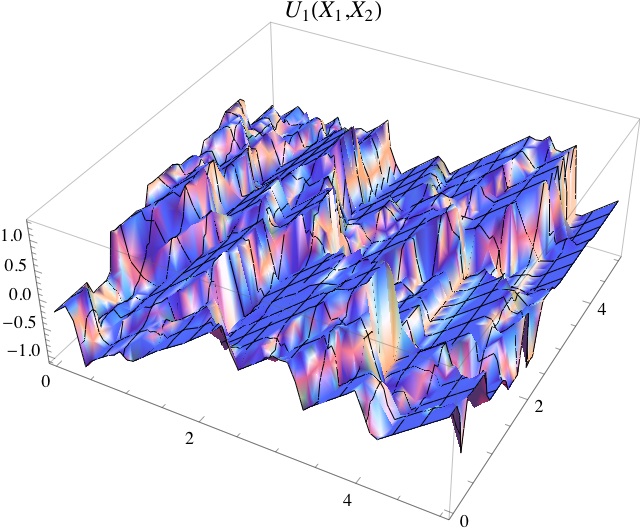}}
\end{center}
\caption{(Color online)
The Lagrangian coordinate $Q_1(\bX)$ (upper panel) and the velocity $U_1(\bX)$
(lower panel) for a realization of the case $n=0$. The coordinate $Q_1$ is monotonically
increasing along direction $X_1$ while $U_1$ shows ramps of slope unity separated by
downward jumps.}
\label{figq1u1_2d}
\end{figure}

Finally we show in Fig.~\ref{figq1u1_2d} the Lagrangian coordinate $Q_1$ (upper panel)
and the velocity component $U_1$ (lower panel) over the Eulerian $\bX$-plane, for the case
$n=0$. 
As explained above, within each of the ``Voronoi-like'' cells shown in Fig.~\ref{figvor_2d}
the Lagrangian coordinate $\bQ$ is constant, so that the surface $Q_1(\bX)$ shows
a series of flat plateaus delimited by the boundaries of these cells, where finite jumps
take place. Along the $X_2$ direction jumps in $Q_{1}$ can be both positive and negative,
with an even distribution as the system is statistically homogeneous and isotropic, but
jumps are always positive along the $X_1$ direction. This is a consequences of elementary
properties of the Legendre transform: as can be seen from the first Eq.(\ref{qx_xq}) and the
fact that $H(\bx)$ is convex, we have the two relations (and similarly in higher dimensions)
\beq
\frac{\pl q_1}{\pl x_1} \geq 0 , \;\;\; \frac{\pl q_2}{\pl x_2} \geq 0 ,
\label{qxmonotonic}
\eeq
which generalize the same one-dimensional property. However, as explained above,
while in one dimension this non-crossing property is sufficient to reconstruct the
direct Lagrangian map, $\bq\mapsto\bx$, in higher dimensions this requires the use
of the second Legendre transform (\ref{phi_convexhull}). Note that by the same convexity
argument we also have
\beq
\frac{\pl x_1}{\pl q_1} \geq 0 , \;\;\; \frac{\pl x_2}{\pl q_2} \geq 0 .
\label{xqmonotonic}
\eeq
Over the ``Voronoi-like'' cells, $\bq$ being constant we can see from Eq.(\ref{vxv0q})
that the velocity components are affine functions of $\bx$. More precisely, we have
for instance $u_1(\bx)=(x_1-q_1)/t$, so that within each cell $u_1$ is constant along
the $x_2$ direction while it grows with a slope $1/t$ along the $x_1$ direction,
as can be checked in the lower panel of Fig.~\ref{figq1u1_2d}. At the boundaries
of the cells, $u_1$ shows positive and negative jumps, with an even distribution,
along the $x_2$ direction, and only negative jumps along the $x_1$ direction,
as can be seen from the behavior of $q_1(\bx)$.

\subsection{Merging and fragmentation}
\label{Merging-fragmentation}

\begin{figure*}
\begin{center}
\epsfxsize=4.4 cm \epsfysize=4.4 cm {\epsfbox{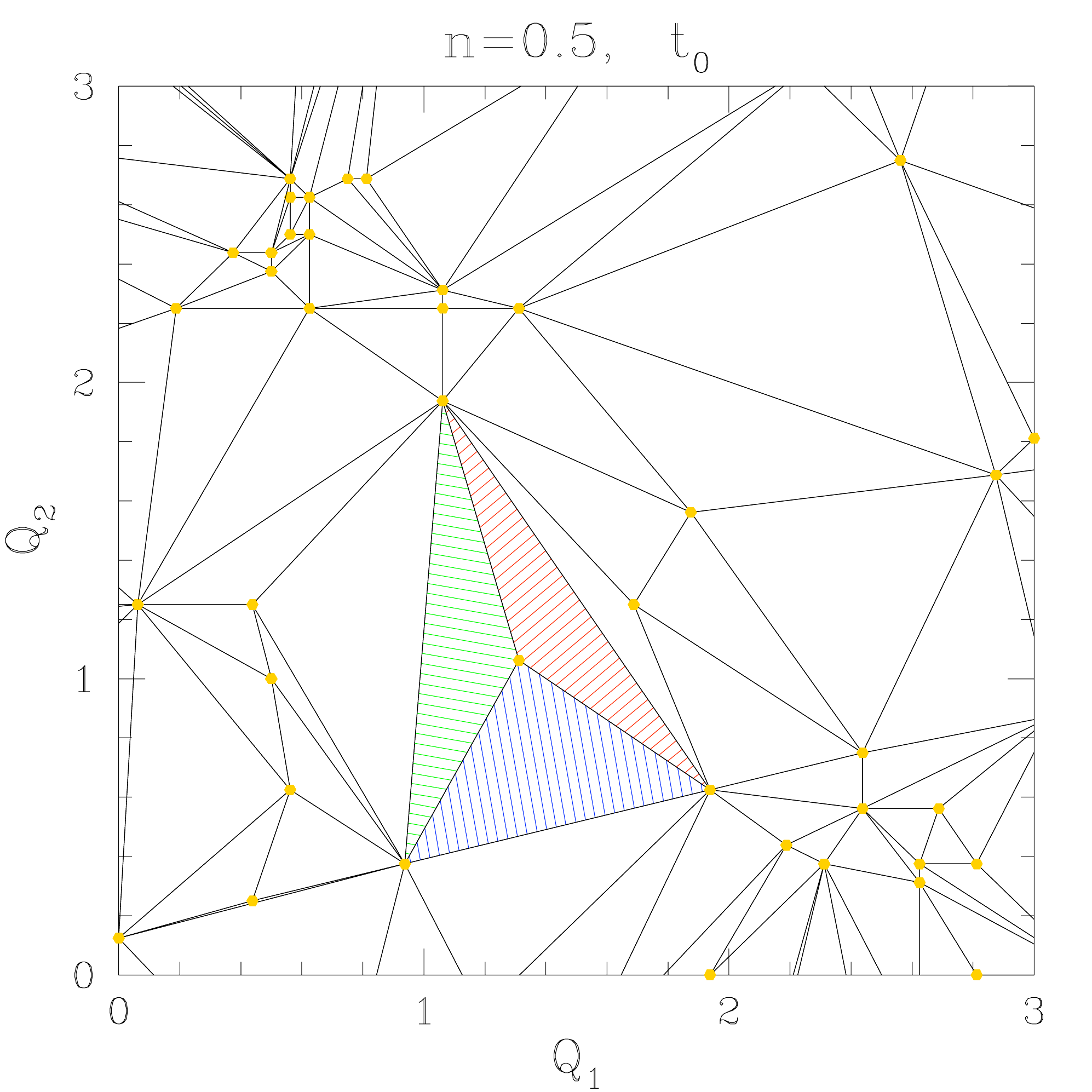}}
\epsfxsize=4.4 cm \epsfysize=4.4 cm {\epsfbox{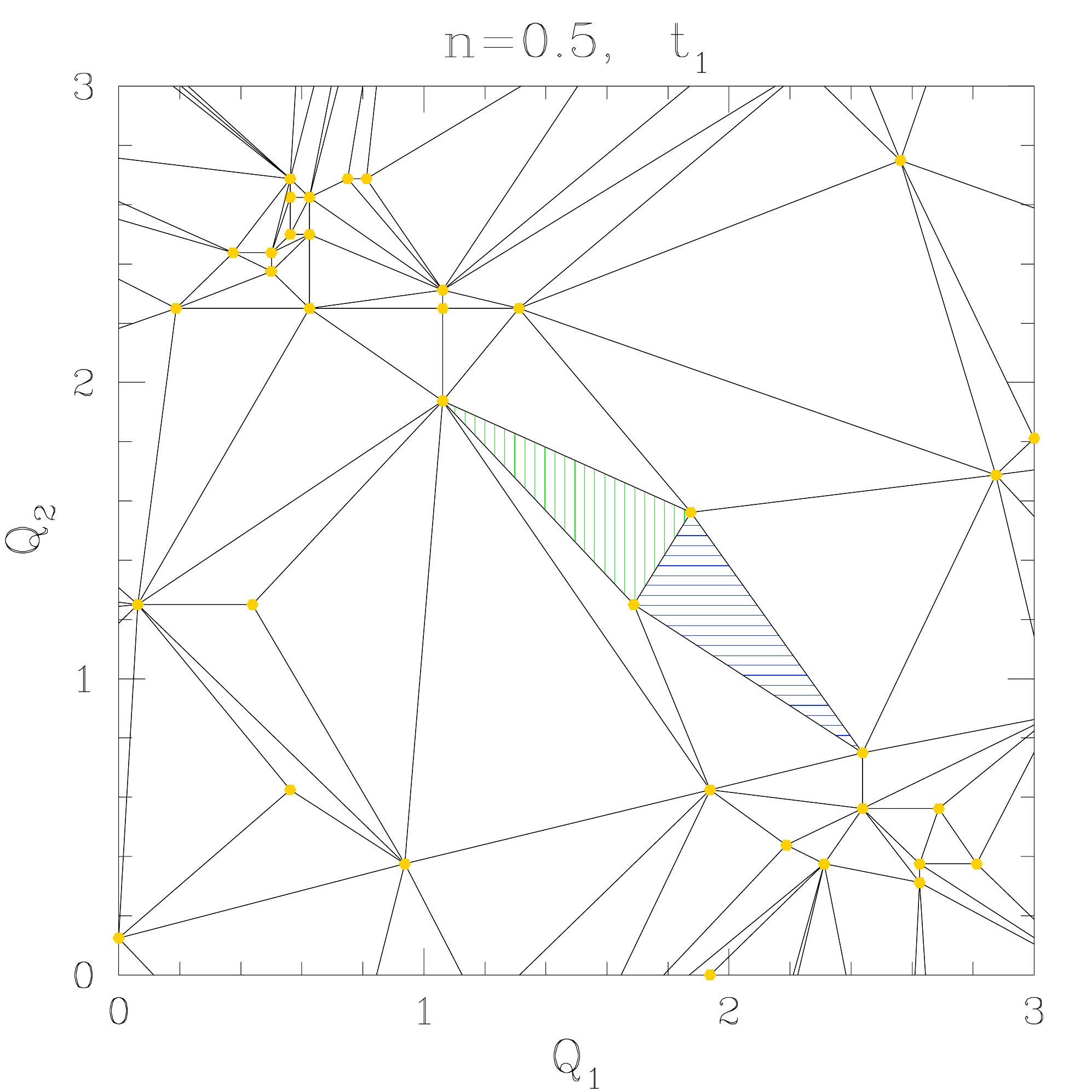}}
\epsfxsize=4.4 cm \epsfysize=4.4 cm {\epsfbox{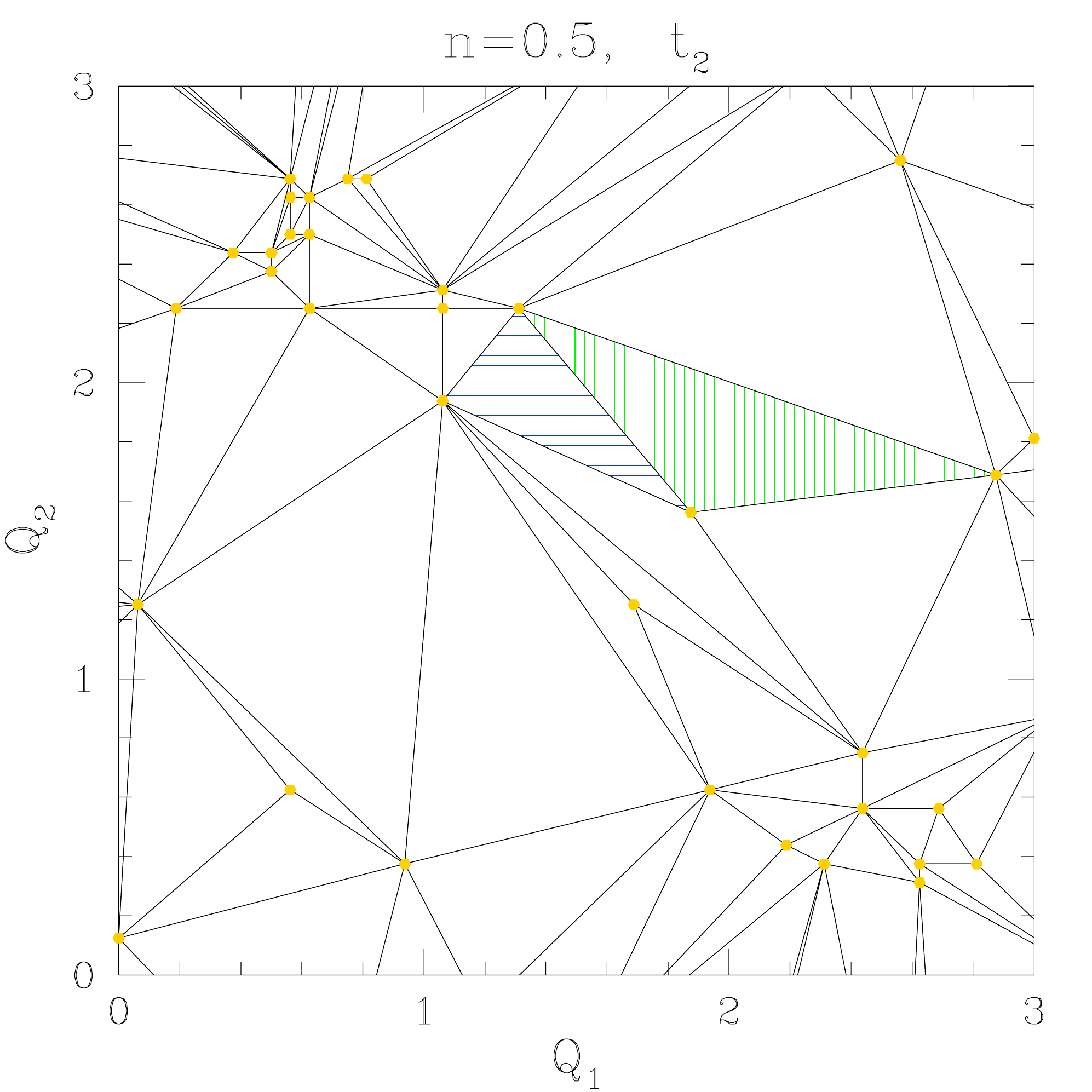}}
\epsfxsize=4.4 cm \epsfysize=4.4 cm {\epsfbox{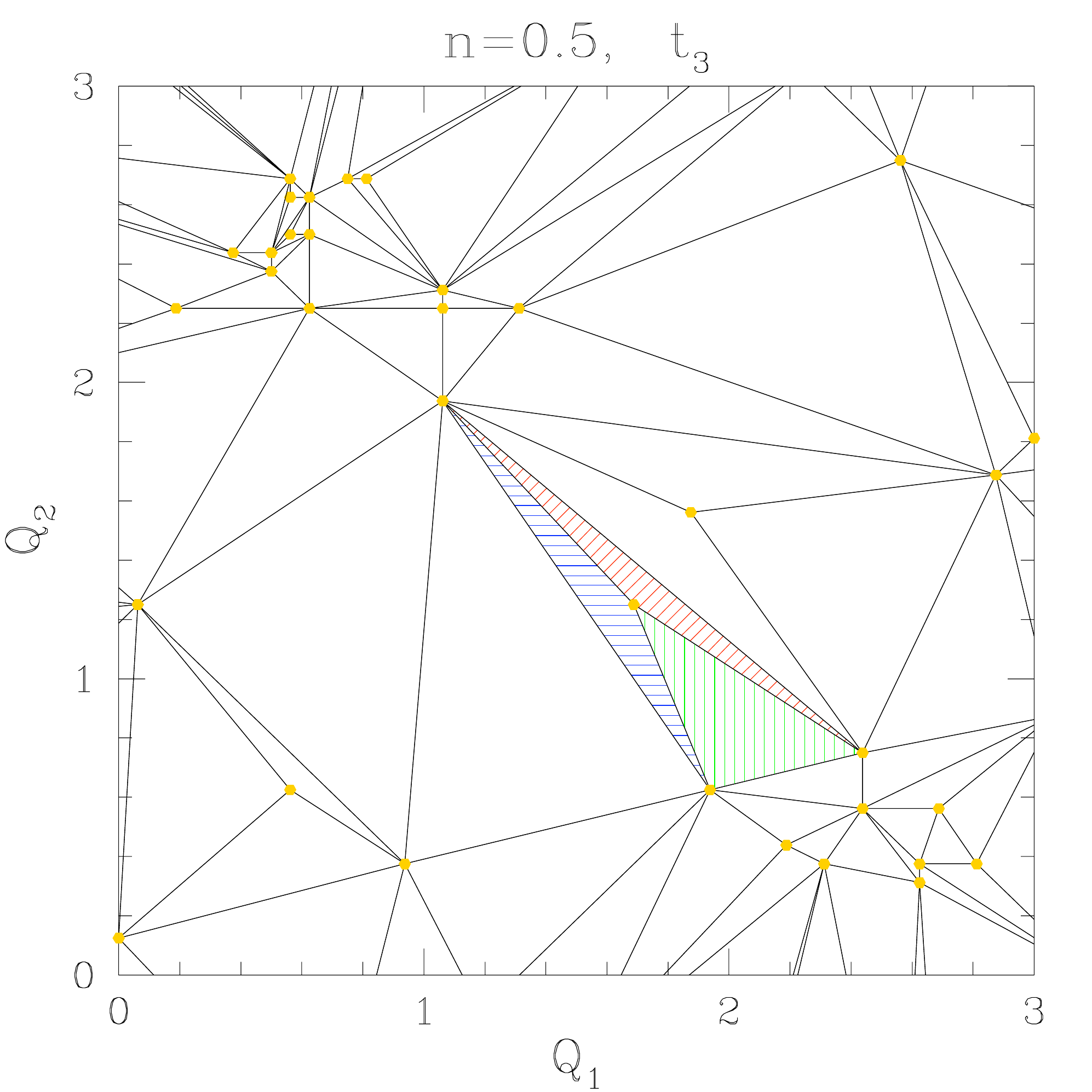}}\\
\epsfxsize=4.4 cm \epsfysize=4.4 cm {\epsfbox{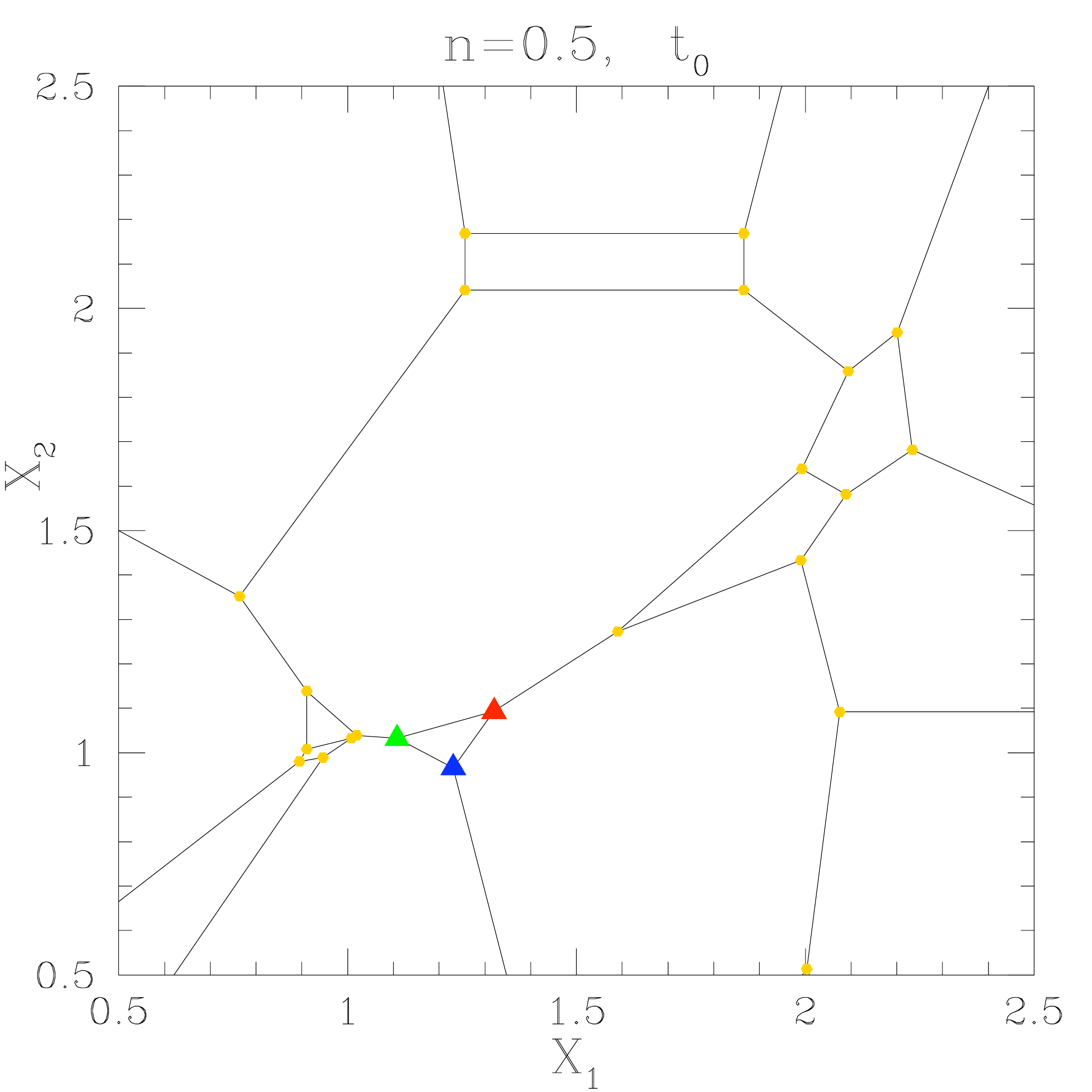}}
\epsfxsize=4.4 cm \epsfysize=4.4 cm {\epsfbox{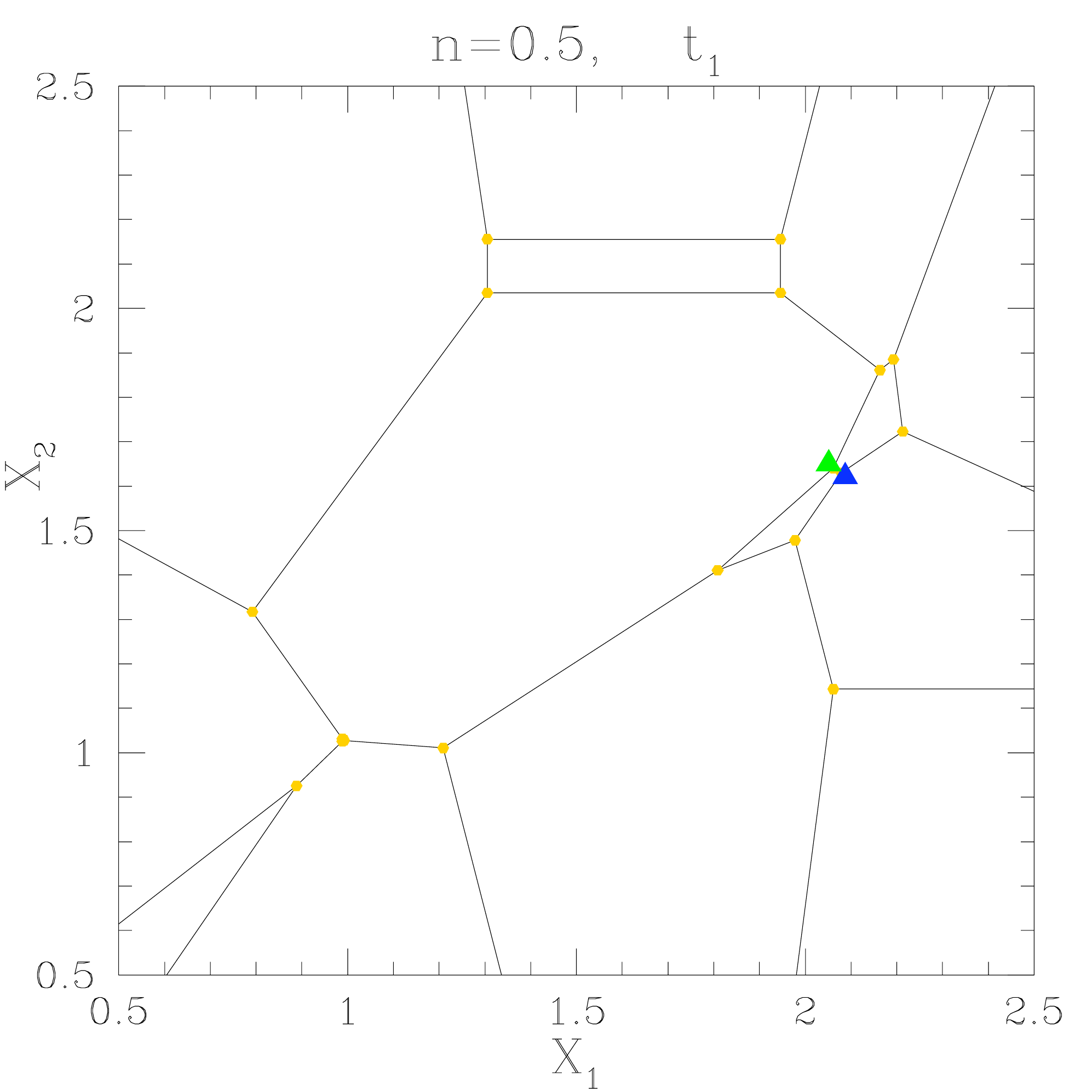}}
\epsfxsize=4.4 cm \epsfysize=4.4 cm {\epsfbox{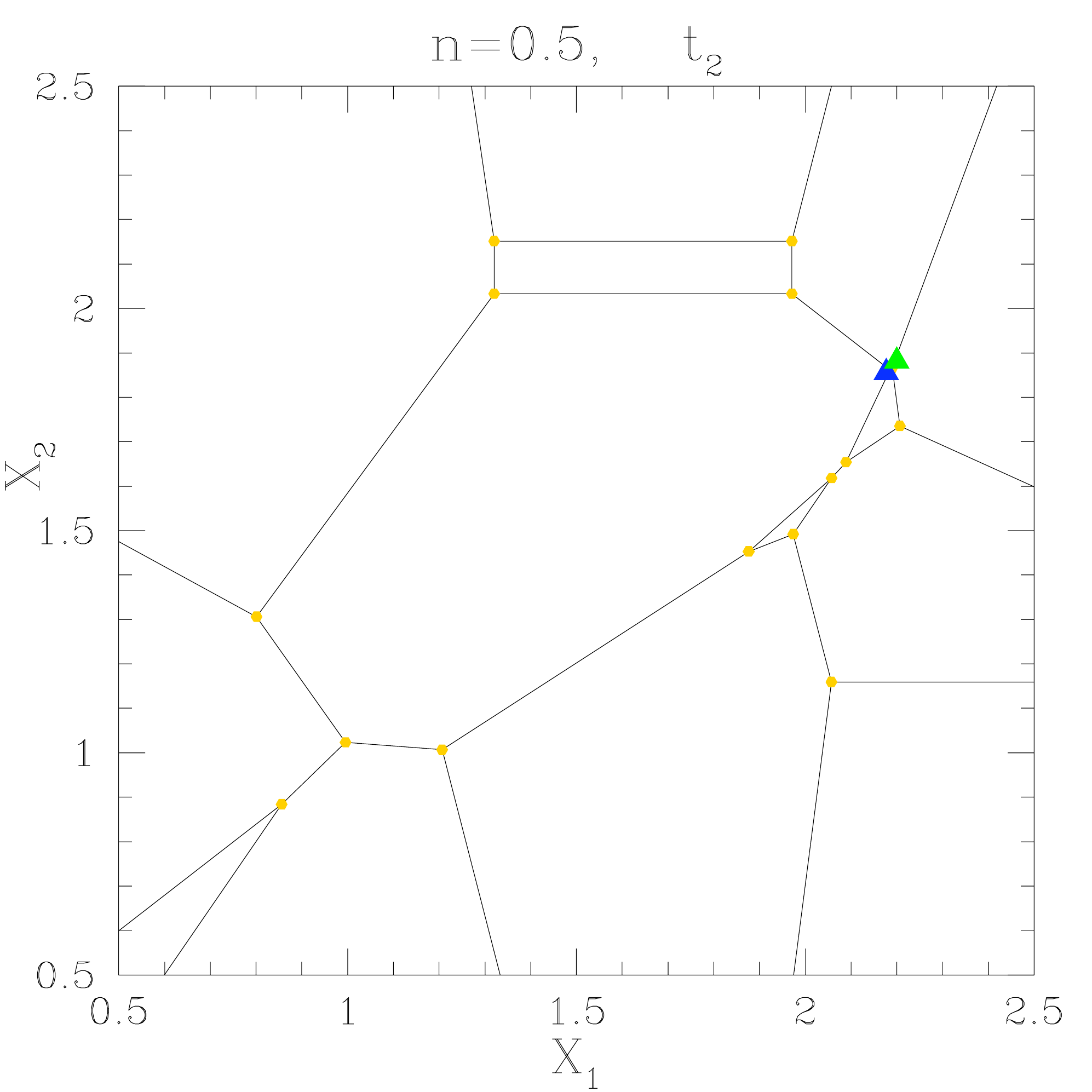}}
\epsfxsize=4.4 cm \epsfysize=4.4 cm {\epsfbox{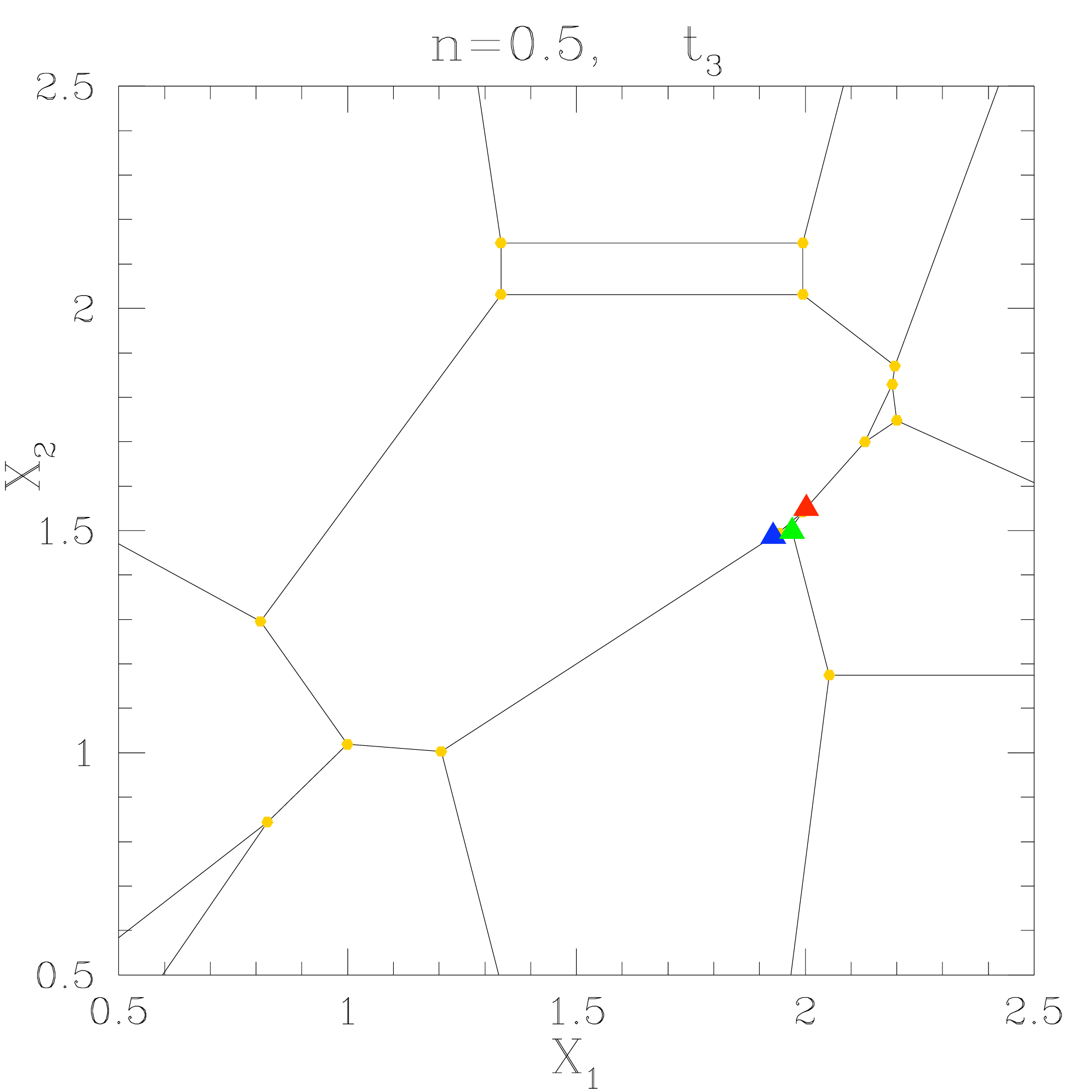}}
\end{center}
\caption{(Color online)
The evolution with time of the Lagrangian (upper row) and Eulerian (lower row)
tessellations, for one field realization with $n=0.5$. We show snapshots obtained
at four successive times $t_0<t_1<t_2<t_3$, from left to right.\\
$\bullet$ At time $t_0$ (first column) we can see in the upper panel three colored triangles
which happen to form a larger triangle. They correspond to the three shock nodes
around $(X_1 \simeq 1.2,X_2 \simeq 1)$ in the lower panel, which form a triangular
``Voronoi'' cell. As seen in the second column, by time $t_1$ these three Lagrangian
triangles have merged to leave the unique larger triangle. In doing so the central vertex
$\bq_c$ has been removed, while in Eulerian space the three shock nodes have merged. \\
$\bullet$ At time $t_1$ we have two colored triangles associated with the two shock nodes
around $(X_1 \simeq 2,X_2 \simeq 1.6)$ that are moving closer (compare with their
position at $t_0$). Between time $t_{1}$ and $t_{2}$ these two shock nodes collide and 
form two new shocks moving outward in a roughly orthogonal direction, as seen in the
third column, while in Lagrangian space there has been a flip.  \\
$\bullet$ A similar process occurs between times $t_2$ and $t_3$, with the collision 
and ``bouncing'' of the shock nodes around $(X_1 \simeq 2.2,X_2 \simeq 1.8)$.\\
$\bullet$ The new triangulation obtained between times $t_1$ and $t_2$ has produced
the configuration with the three colored triangles shown at time $t_3$, which now
form a unique larger triangle (that was not the case at the earliest time $t_0$).
They are associated with the three shock nodes around
$(X_1 \simeq 2,X_2 \simeq 1.5)$ which are moving closer. Thus, we recover a central
configuration similar to the one shown at time $t_0$, and these three Lagrangian
triangles and Eulerian shock nodes merge at a later time $t_4$ (not shown).
}
\label{figtime_2d}
\end{figure*}

As time goes on, the tessellation and its associated triangulation are bound to evolve.
For any dimension, Lagrangian vertices (i.e. contact points between the
linear Lagrangian potential and its convex hull) at a time $t_2$ form a subset of the
vertices obtained at an earlier time $t_1<t_2$, as can be most easily seen from
the paraboloid construction (\ref{Paraboladef}).
However, it is clear that a naive merging of neighboring triangles in the
Lagrangian-space triangulation shown in Fig.~\ref{figtriang_2d} does not always
produce a new triangulation, as four-sided polygons would usually appear.
Therefore, new triangulations obtained at time $t_2>t_1$ cannot be merely coarser
partitions of the initial one. That is, even though
all vertices seen at $t_2$ were already vertices at $t_1$, triangles seen at $t_2$
are not always the union of smaller triangles obtained at $t_1$.
This implies that a redistribution of matter throughout the triangulation must take place.
That can only be so through fragmentations of shock nodes.

In order to show how it works let us simply examine, as in Fig.~\ref{figtime_2d},
successive snapshots of the triangulations/tessellations. Those presented have been 
obtained for one realization of a field of index $n=0.5$ (this value was chosen so 
that cells are of similar sizes to make
the figure easier to read but the qualitative behavior is identical to the $n=0$ case). 
We display in the upper row the Lagrangian space triangulations and in the
lower row the Eulerian space ``Voronoi-like'' diagrams, obtained at four successive
time (from left to right). The time steps have been chosen so that only a few
rearrangements can be seen from one snapshot to the other.  For each output, we color 
the triangles (in Lagrangian space), and the associated
shock nodes (shown by triangular symbols in the lower panel in Eulerian space),
that are going to be affected. Note that the relative orders of the
Lagrangian triangles and of the Eulerian nodes match, in agreement with the
monotonicity relations (\ref{qxmonotonic}) and (\ref{xqmonotonic}).

On these successive snapshots, only two types of triangle rearrangement can be observed:
\begin{itemize}
\item a \textsl{flip} ($2\rightarrow 2$): diagonal inversion in a quadrangle formed by
two adjacent triangles;
\item a \textsl{3-merging} ($3\rightarrow 1$): merging of three adjacent triangles into
a single one.
\end{itemize}

These properties can be understood from the convex hull construction. For a generic time $t$,
the convex hull is entirely made of triangular facets. This means that no four summits of the
convex hull are coplanar. As time is changing, critical time values can be reached where
four points become planar. In between two critical time values the Lagrangian
triangulation simply does not evolve while the associated Eulerian tessellation evolves
continuously. At critical times, there are two possibilities: either 
one of the four points is within  the triangle formed by the three others or not. The latter
case corresponds to a flip transition; the former to a 3-merging.
Here we must note that the existence of ``flip events'', that is $2\rightarrow 2$ collisions
with a redistribution of matter, was already noticed in \cite{Gurbatov1991}
(the two-clump collision accompanied by mass exchange shown in their Fig.~6.21).

Other transitions corresponding to a higher-order degeneracy (5 or more coplanar points)
are possible in principle but have a vanishing probability to occur.

Thus, we now clearly see how the system evolves with time through a
succession of two-body and three-body collisions.
Two-body collisions of shock nodes only redistribute matter over two new shock nodes
and redefine the Lagrangian triangulation, which allows the formation of specific
central configurations where the union of three neighboring triangles forms a
larger triangle. Next, such configurations allow the removal of the central
Lagrangian vertex, through the merging of the three triangles, which corresponds to
a simultaneous merging of three shock nodes in Eulerian space.

\begin{figure}
\begin{center}
\epsfysize=7. cm {\epsfbox{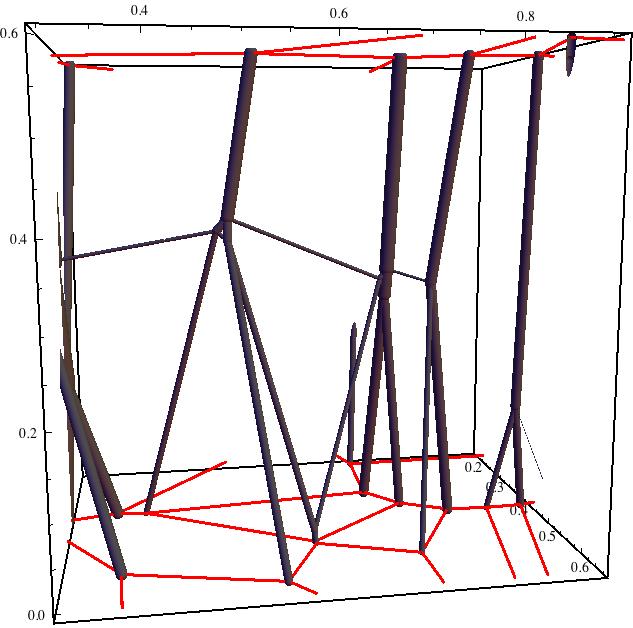}}
\end{center}
\caption{(Color online)
A merging-fragmentation tree obtained for one realization for the case $n=0$, $d=2$.
The bottom plane corresponds to the initial time where the Voronoi-like structure showing
the location of the halo nodes is shown. The top plane corresponds to the final time. The
tubes in between show the halo trajectories. The section of each tube is proportional to its
corresponding halo mass. One can observe on this picture examples of $2\leftrightarrow  2$
halo diffusions and $3\to 1$ mergings. 
Units are arbitrary. 
}
\label{figtree_2d}
\end{figure}

To conclude this paragraph we present in Fig.~\ref{figtree_2d} a realization of a small fraction
of a merging-fragmentation graph obtained for the case $n=0$. The horizontal plane
corresponds to the Eulerian $\bx$-plane while the vertical axis is the time (units are
arbitrary). Tubes correspond to trajectories of halo nodes and their sections are proportional
to their mass.
One can see that, similarly to the 1D case, the trajectories are straight lines in between
the collisions taking place at critical times. In the $d=2$ case, collisions lead either to
$2\leftrightarrow  2$ scatterings with mass exchange or to $3\to 1$ mergings. The former
process is unknown in $d=1$ case. The net result of these effects \footnote{It can also be
remarked that critical time values are strongly correlated, as in the center of the figure few
subsequent fragmentations/mergings happen on a rather short time scale as compared to the
typical time scale of this figure. We did not try to investigate in more details the time
correlations of these critical  events.} is to progressively diminish the number density of
halos -- and to augment their average mass -- in agreement with the scaling law (\ref{Lt}).

\subsection{Velocities in fragmentation events}
\label{Velocities}

\begin{figure*}
\begin{center}
\epsfxsize=4.5 cm \epsfysize=4.5 cm {\epsfbox{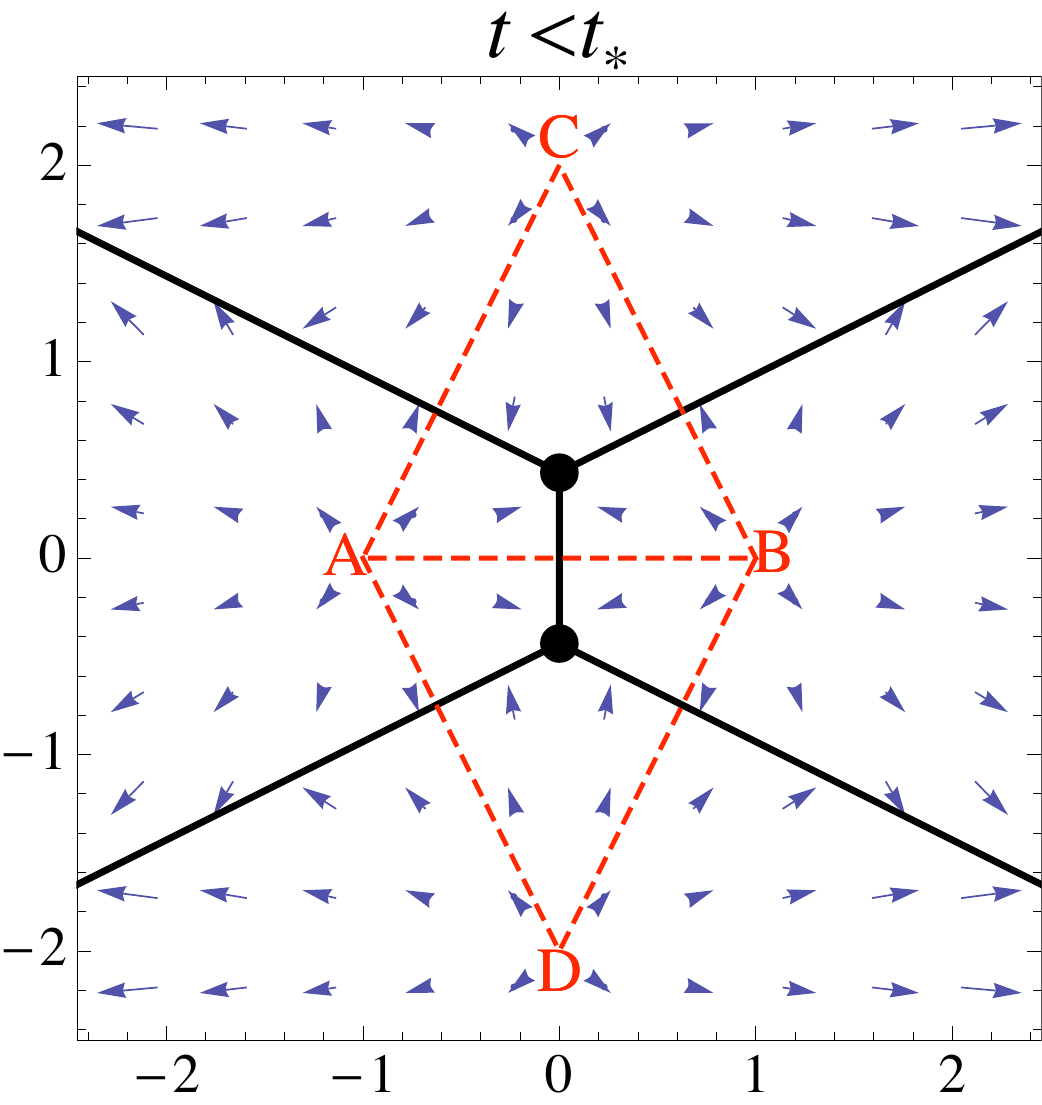}}
\epsfxsize=4.5 cm \epsfysize=4.5 cm {\epsfbox{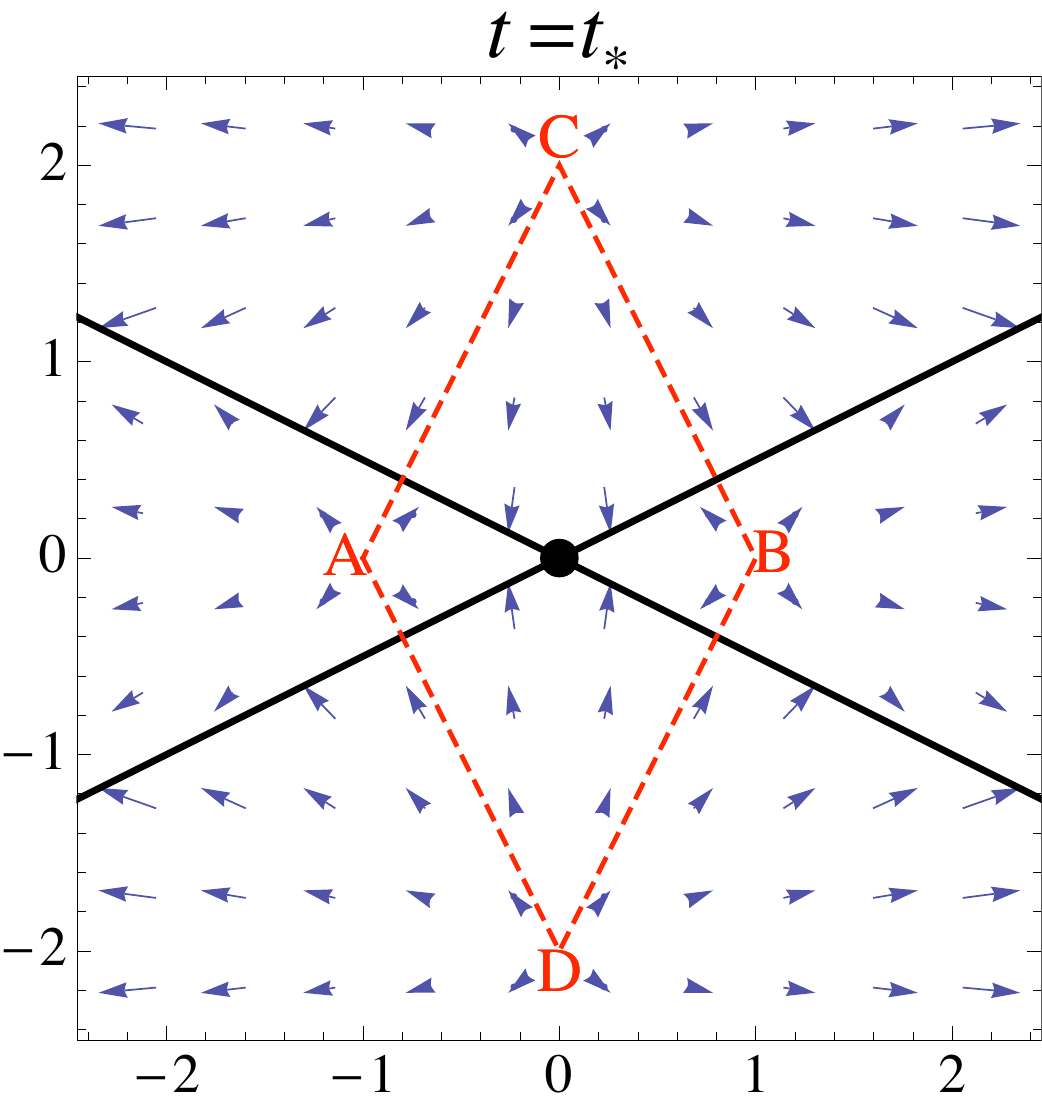}}
\epsfxsize=4.5 cm \epsfysize=4.5 cm {\epsfbox{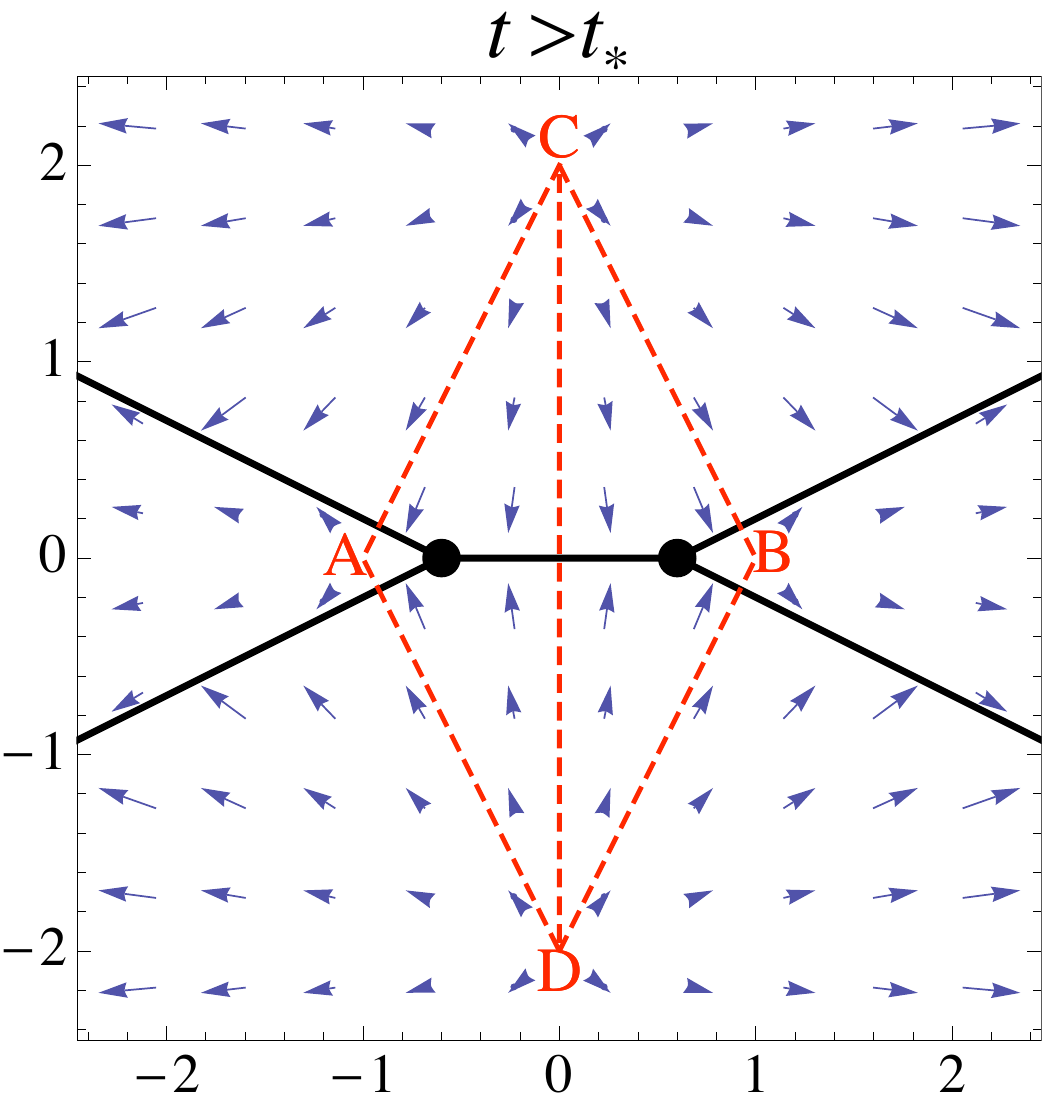}}
\end{center}
\caption{(Color online)
A collision of two shock nodes that gives rise to two new shock nodes with a
redistribution of mass.}
\label{figcoll_2d}
\end{figure*}

Although it was already noticed in \cite{Gurbatov1991}, the existence of flip-like transitions,
that is $2\rightarrow 2$ collisions with mass exchange in 2D, is not widely known in
the cosmological community. In particular, despite the ``geometrical model'', which gives
rise to such events, has been studied in numerical works
\cite{Vergassola1994,1992ApJ...393..437K} in the context of the formation of large-scale
structures in cosmology, other workers in this broader field (who have not necessarily studied
the problematics associated with the Burgers equation) are not always aware of the fact
that this peculiar model implies such splitting events in dimensions two and higher
(at least, the authors of this paper were not aware of this behavior before working on the
present study).

Let us point out that the Eulerian-space and Lagrangian-space tessellations have a
different status in this regard.
Indeed, the Voronoi-like tessellation itself, shown in the bottom row of
Fig.~\ref{figtime_2d} and in Fig.~\ref{figvor_2d}, is an Eulerian-space construction that
describes the regular regions. It is thus entirely defined by the Burgers equation in its
inviscid limit, through the Hopf-Cole solution (\ref{psixpsi0q}). This is not so for the
triangulation that lives in Lagrangian space, shown in the upper row of
Fig.~\ref{figtime_2d} and in Fig.~\ref{figtriang_2d}. 
It requires a prescription on where the matter is actually going, that is, one must
add to the Burgers equation a second equation for the evolution of the density field.
One can choose the ``standard'' continuity equation, or the modified  continuity equation
(\ref{cont1}) associated with the ``geometrical model'' described in Sect.~\ref{Zero-viscosity},
which we study in this paper. Different models lead to different Lagrangian-space
tessellations, since mass clusters may or may not split depending on the chosen
prescription. This also means that, even though they do not change the Voronoi-like diagrams,
that is, the cellular structure built by the shock manifold, these different models put
matter at different positions on this shock manifold. As we have already mentioned,
in the ``standard'' model mass clusters cannot split but they may leave shock nodes,
as found in \cite{Bogaevsky2004,Bogaevsky2006}, whereas in the ``geometrical model''
mass clusters can split but they are always located on shock nodes (and each node
is associated with a mass cluster). Of course, the latter property is due to the
geometrical construction of this model, associated with the Legendre transformations and
convex hull constructions described in Sect.~\ref{Zero-viscosity}. 
As explained in the previous section, and illustrated in Fig.~\ref{figtime_2d},
this geometrical constraint (requiring that the mass distribution is always defined
by the dual Eulerian Voronoi-like tessellation/Lagrangian triangulation) implies in
dimensions 2 and higher that collisions can redistribute matter among mass clusters.
In contrast, in the ``standard'' model, where mass clusters do not split, the tessellations
are not sufficient to recover the matter distribution since clusters are not necessarily
located at the summits of the Voronoi-like cells (then, one needs to integrate the continuity
equation over previous times to solve the problem).

In order to illustrate more clearly how the $2\rightarrow 2$ collisions proceed 
from a dynamical point of view, we show a simple example of such a flip-like
transition in Fig.~\ref{figcoll_2d}. Such an event was also described in Fig.~6.21
of Ref.~\cite{Gurbatov1991}, but we complete the picture by plotting the associated
Eulerian velocity field $\bu(\bx,t)$ and discussing in somewhat more details how this
proceeds. This example also allows us to compare in Sect.~\ref{comparison} below
the behavior of the ``geometrical model'' with that of the ``standard'' model,
using the results obtained recently by \cite{Bogaevsky2004,Bogaevsky2006}.
We consider the simplest case where the initial velocity
potential $\psi_0(\bq)$ shows four spikes, labeled as points $A,B,C$ and $D$,
of coordinates
\beqa
L>\ell>0 & : & A= (-\ell, 0) , \;\; B= (\ell,0) , \nonumber \\
&& C=(0,L) , \;\; D=(0,-L) ,
\label{ABCD}
\eeqa
with
\beq
\psi_0(A)=\psi_0(B)=0 , \;\; \psi_0(C)=\psi_0(D)= \psi_+>0 .
\label{psiABCD}
\eeq
Other points in the $\bq$-plane have much smaller values of $\psi_0$ so that they
do not play any role at the time of interest (we can take $\psi_0(\bq)=-\infty$ everywhere
outside of points $\{A,B,C,D\}$). This simple case obeys at all times the two parity
symmetries $x_1\leftrightarrow-x_1$ and $x_2\leftrightarrow-x_2$. It is easiest
to analyze the system with the help of the paraboloid construction (\ref{Paraboladef}).
At early time (left panel in Fig.~\ref{figcoll_2d}), paraboloids have a large curvature
(i.e. they are highly peaked)
so that Eulerian positions $\bx$ in the neighborhood of each summit $\{A,B,C,D\}$
are governed by the closest of these four points (i.e. paraboloids of center $\bx$
make first contact with the closest peak among $\{A,B,C,D\}$).
In particular, from Eq.(\ref{vxv0q}) the velocity field $\bu(\bx)$ outflows from each
of these points, as seen in left panel of Fig.~\ref{figcoll_2d} (blue arrows).
Next, the ``domain of influence'' of summit $A$ in Eulerian space (i.e. its ``Voronoi-like''
cell) is delimited by straight lines defined as the
set of points $\bx$ such that the first-contact paraboloid of center $\bx$ simultaneously
touches summits $A$ and $B$, $A$ and $C$, or $A$ and $D$ (and similarly
for other summits). Of course, by symmetry the frontier between the $A$-cell
and the $B$-cell is orthogonal to the segment $(AB)$. These cells are shown by the
black lines in Fig.~\ref{figcoll_2d}. Thus, at early times we have the configuration
displayed in the left panel, with two shock nodes, shown by the big black dots,
at the vertices of these Eulerian cells. Then, the matter that has fallen into the upper
shock node comes from the upper triangle $(ACB)$, whereas the lower shock node
contains the matter from the lower triangle $(ADB)$.
This gives the triangulation of the Lagrangian $\bq$-space, which we show by the
red dashed lines in the figure. Note that in each panel we superpose the Eulerian
space (the velocity field marked by arrows, the ``Voronoi-like'' cells marked by
solid lines and the shock nodes marked by big dots) and the Lagrangian space
(the triangulation marked by the dashed lines).
From these partitions of the $\bx-$ and $\bq-$spaces we can also read the
convex functions $H(\bx)$ and $\varphi(\bq)$ (i.e. this gives the projection of their
planar facets).

As time grows the two shock nodes move closer towards the center of the figure,
until the merge at time $t_*$ shown in the middle panel. At this time, the
paraboloid of center $(0,0)$ makes simultaneous contact with the four summits
$\{A,B,C,D\}$, and the two triangles obtained at earlier times in the left panel
merge to form a losange. 

Next, at later times we obtain the configuration shown in the right panel,
such that the paraboloid of center $(0,0)$ only makes contact with summits
$C$ and $D$. We now have two shock nodes moving outward from the center along the
horizontal axis. The Lagrangian triangle associated with the left (resp. right) shock node
is now $(ACD)$ (resp. $(BCD)$). Therefore, the matter within the losange $(ACBD)$
has been redistributed. In particular, the triangle $(ACB)$ of the left panel has been
split into two parts, the left half going into the left shock and the right half going
into the right shock. Thus, the unique central shock node obtained at time $t_*$
has fragmented into two new shock nodes and some particles that had coalesced at
earlier times $t<t_*$ have separated and now belong to two different objects.

It is interesting to note that it had been noticed in numerical simulations
\cite{1992ApJ...393..437K} that massive clumps seen in $N$-body simulations of the
gravitational dynamics generally are associated with several nodes in the corresponding
adhesion-model simulations (i.e. defined by the same initial conditions).
This was interpreted as an inaccurate description of the merging process, as the
Burgers dynamics does not take into account local gravitational forces. Our results
show that such differences may also be due to the specific fragmentation process
associated with this Burgers dynamics.

\subsection{Comparison with the ``standard'' model}
\label{comparison}

The peculiar fragmentation process described in the previous section arises from
our prescription (\ref{rhoHvarphi}) for the density field, that is, from the modified
continuity equation (\ref{cont1}). If we choose to set the right hand side of Eq.(\ref{cont1})
to zero, that is, for finite $\nu$ the density field obeys the standard continuity equation,
we would obtain a different result with no fragmentation in the inviscid limit
as studied in  \cite{Bogaevsky2004,Bogaevsky2006}.
To illustrate this point, let us consider the configuration shown in Fig.~\ref{figcoll_2d}
for a small non-zero viscosity, $\nu>0$. From Eqs.(\ref{Hopf1})
and (\ref{psixpsi0q}) we know that in the limit $\nu\rightarrow 0^+$ the velocity
field at any time converges towards the inviscid solution displayed in
Fig.~\ref{figcoll_2d}. The main difference is simply that at finite viscosity the
discontinuities along the shock lines (black solid lines in the figure) are 
smoothed over a small finite distance. In particular, the central shock node obtained
at time $t_*$ (middle panel) has a small nonzero extension. Then, since the
parity symmetries of the system are still exactly satisfied, we can see that at
infinitesimally later times the particles located in the left part of this finite-size
cloud experience a local velocity field that points towards the left, whereas 
particles located in the right part see a velocity field that points towards the right.
However, if we use the standard continuity equation this does not lead to a splitting into
two new symmetrical halos, moving further apart as time grows.
Indeed, the differences between
the velocities $\bu(\bx,t)$  of the left and right parts of the small central cloud go to
zero with $\nu$ (along with the size of the cloud) so that the small cloud stays in the
middle of the new horizontal shock line in the inviscid limit. 
On the other hand, if the spikes of the initial potential $\psi_0$ at points $\{A,B,C,D\}$
are not infinitesimally thin, two new horizontal nodes appear as in the figure and are fed
by the matter flowing from the regular parts of these spikes (so that one obtains three
mass clusters, a motionless central one which has stopped growing and two small
outward-going ones which have just formed).

As mentioned earlier, the evolution of matter within shock lines was studied in
\cite{Bogaevsky2004,Bogaevsky2006}, using the standard continuity equation
for the evolution of the density field. It was found that the trajectories obtained in the
inviscid limit exist and are unique, so that trajectories that pass through a point at a given
time coincide at all later times. Therefore, there is no fragmentation of halos in this
prescription.
On the other hand, clusters can stop growing and leave the shock nodes (while remaining
on shock lines), in agreement with the discussion above where we explained how within
this prescription a small halo would remain motionless at the center of the right panel
of Fig.~\ref{figcoll_2d}. 
This approach can be extended to more general Hamilton-Jacobi equations, with convex
Hamiltonians. Then, one again obtains a unique coalescing flow 
\cite{Bogaevsky2006,Khanin2010}.

The evolution with time of the shock lines themselves (bifurcations, transitions) in 2D
and 3D was studied in details in \cite{Bogaevsky2002}, for general initial conditions.
In particular, both the $2\rightarrow 2$ ``flip'' and $3\rightarrow 1$ merging
events described above in the ``late-time'' regime for the 2D case correspond to
the $A_1^4$ space-time event of Fig.2 in \cite{Bogaevsky2002}, where the
first-contact paraboloid makes simultaneous contact with four ``non-degenerate''
points (in this classification a contact point is called ``non-degenerate'' if 
$(\cP_{\bx,c}-\psi_0)$ has a non-zero second derivative there -- in our case the second
derivative is actually infinite).

\subsection{Momentum exchange through shock node collisions}
\label{Momentum-2d}

In the regime studied in this article, associated with the power-law initial conditions
(\ref{ndef}) or the late-time stage, not only is all the matter content distributed within shock
nodes but for $d\ge 2$ shocks form a unique connected set that spans the entire space. As a 
consequence, conservation of the ``effective momentum'' defined in Eq.(\ref{Idef}) during
node collisions cannot be inferred  from the general result described in 
sect.~\ref{Momentum}: no volume $\cV_{\bx}$ with a regular 
boundary can be drawn around a finite number of nodes. 
We explore here how the actual momentum of the particles
can be exchanged  through shock node collisions.

First, let us express the momentum of a given shock node. We note $\{A,B,C\}$ the
summits of its associated triangle in Lagrangian space, which corresponds to a triangular
facet of the convex hull $\varphi(\bq)$ of the Lagrangian potential. We consider time values
between critical times, that is, a time interval during which the triangle $\{A,B,C\}$
is left unchanged.

The relation (\ref{VorBoundaries}) can be written for $A$ and $B$ and for $A$ and $C$. 
Differentiating those equations with respect to time $t$, we get the node velocity $\bv$
along the 
vectors $\bq_{AB}\equiv\bq_{B}-\bq_{A}$ and $\bq_{AC}\equiv\bq_{C}-\bq_{A}$,
\begin{eqnarray}
\bv\cdot\bq_{AB}&=&\psi_{0}(\bq_{A})-\psi_{0}(\bq_{B});\\
\bv\cdot\bq_{AC}&=&\psi_{0}(\bq_{A})-\psi_{0}(\bq_{C}).
\end{eqnarray}
Furthermore the mass of the node is given by the area of the triangle
\beq
m = \frac{\rho_0}{2} | \, \bq_{AB} \times \bq_{AC} | \,.
\label{trianglemass}
\eeq
Introducing the momentum $\bp=m\bv$, those results can be recapped in a single
expression. Defining the points $\{\cA,\cB,\cC\}$ in a fictitious 3D space
\beq
\cA = \left\{q_{1A},\  q_{2A},\  \psi_0(\bq_A) \right\} , \dots ,
\label{cAdef}
\eeq
we obtain
\beq
\cP \equiv \left\{\bp,\  m \right\} = \frac{\rho_0}{2}\ \vec{\cA\cB}\times\vec{\cA\cC} ,
\label{pshock}
\eeq
where $\vec{\cA\cB}=\cB-\cA$, provided $\{\bq_{AB},\bq_{AC}\}$ is positively oriented.

Note that we use the letter ``$\bv$'' for the shock node velocity to distinguish this quantity
from the Eulerian velocity $\bu(\bx)$, which is discontinuous along shock lines.
Moreover, as seen in the previous section and in Fig.~\ref{figcoll_2d}, in dimensions
$d\geq 2$ the former is not given by the mean of ``left'' and ``right''  Eulerian velocities
(at $t>t_*$ the two shock nodes shown in the right panel of Fig.~\ref{figcoll_2d} have
constant finite outward velocities $\bv$ whereas the horizontal component of $\bu$
is linear over $x_1$ and goes to zero at the center of the figure). Thus, we consider in this
section the standard momentum of the particles, rather than the ``effective momentum''
introduced in Sect.~\ref{Momentum}.

Let us now consider a $2\rightarrow 2$ shock collision, such as the one shown in
Fig.~\ref{figcoll_2d}. Without loss of generality, we can still choose the common
edge $(AB)$ before collision on the horizontal axis, $C$ in the upper half plane and
$D$ in the lower half plane in Lagrangian space, but with otherwise arbitrary
coordinates (i.e. the triangles need not be symmetrical). 
Then, before collision the total 3-momentum reads as
\beqa
\cP &=&  \{\bp_{ABC} + \bp_{ADB},\  m_{ABC}+m_{ADB}\}\nonumber\\
&=&\cP_{ABC} + \cP_{ADB} ,
\eeqa
where we introduced the 3-momenta of the two shock nodes, associated with the
Lagrangian triangles $(ABC)$ and $(ADB)$. At the time of the collision, 
paying attention to the orientation of the Lagrangian-space triangles, we obtain
from Eq.(\ref{pshock}),
\beqa
\cP& =  &\frac{\rho_0}{2} 
\left[ \vec{\cA\cB}\times\vec{\cA\cC} + \vec{\cA\cD}\times\vec{\cA\cB} \right] \nonumber \\
& = & \frac{\rho_0}{2} \vec{\cA\cB}\times\vec{\cD\cC} \nonumber \\
& = & \frac{\rho_0}{2} \left[ \vec{\cD\cC}\times\vec{\cD\cA} + \vec{\cD\cB}\times\vec{\cD\cC} \right]
\nonumber \\
& = & \cP_{DCA} + \cP_{DBC} .
\eeqa
At collision time, we see clearly that both the mass and the 2-momentum $\bp$ are
conserved by the collision, as the 3-momenta associated with the two new shock nodes
again sum up to $\cP$.
Note that the fact that both triangles $(DCA)$ and $(DBC)$ are counterclockwise
(so that Eq.(\ref{pshock}) applies with no further negative sign) comes from the
constraint that we have a $2\rightarrow 2$ collision rather than a $3\rightarrow 1$
event (i.e. the 2D segment $[CD]$ intersects the segment $[AB]$).

Finally, let us consider a $3\rightarrow 1$ merging event, such as the one shown in
left panel of Fig.~\ref{figtime_2d}.
Thus, we choose a direct triangle $(BCD)$ with an inner summit $A$.
Before merging the 3-momenta associated with the three shock nodes (and the
three Lagrangian-space triangles) read as
\beqa
&& \cP_{ABC} = \frac{\rho_0}{2} \vec{\cA\cB}\times\vec{\cA\cC} ,  \;\;
\cP_{ACD} = \frac{\rho_0}{2} \vec{\cA\cC}\times\vec{\cA\cD} , \nonumber \\
&& \cP_{ADB} = \frac{\rho_0}{2} \vec{\cA\cD}\times\vec{\cA\cB},
\eeqa
whence after straightforward manipulations
\beq
\cP_{ABC} + \cP_{ACD} + \cP_{ADB} = \frac{\rho_0}{2} \vec{\cB\cC}\times\vec{\cB\cD}
= \cP_{BCD} .
\eeq
Therefore, momentum and mass are again conserved by the $3\rightarrow 1$ collision.

The analysis described above, based on expression (\ref{pshock}), shows
that momentum is conserved in the ``late-time regime'' by shock node collisions.
Note that this only holds at a {\it global} level: the total momentum of the two nodes
is conserved in a fragmentation event but the momentum of a given node is not
necessarily equal to the total initial momentum of  the matter particles it contains.
Thus, in-between arbitrary times $t_1<t_2$, momentum is only conserved over disjoint
sets of particles, such that particles in a given set have only been in contact during this
time interval with particles of the same set (so that there has been no exchange of
momentum between different groups).

In the $d=1$ case though, there are only merging events and momentum conservation
is therefore ensured at the {\it local} level, in agreement with well-known general
results \cite{Burgersbook}.

It is interesting to note that the momentum (\ref{pshock}) coincides with
Eq.(\ref{I1psi0}) if the initial velocity potential $\psi_0$ is affine over the three edges
$(AB)$, $(BC)$ and $(CA)$ of the Lagrangian-space triangle $\{A,B,C\}$.
In particular, this means that the momentum of a given node is equal to
the total initial momentum of  the matter particles it contains if the initial velocity
potential $\psi_0(\bq)$ is affine over this triangle $\{A,B,C\}$ (this also explicitly
shows that both quantities are generically different for arbitrary potential $\psi_0$).
Then, for the power-law initial conditions (\ref{ndef}) where the transition
to the ``late-time'' regime takes place at an infinitesimally small time,
$t_*\rightarrow 0^+$, with a characteristic scale $L(t_*)$ of the Lagrangian-space
triangulation that goes to zero, we can see that the initial momentum is conserved
(in a global sense) since the piecewise affine approximation of $\psi_0$ defined
by this triangulation  converges to $\psi_0$.

\section{Three-dimensional dynamics and beyond}
\label{Three-dimension}

\begin{figure*}
\begin{center}
\epsfxsize=5 cm \epsfysize=5 cm {\epsfbox{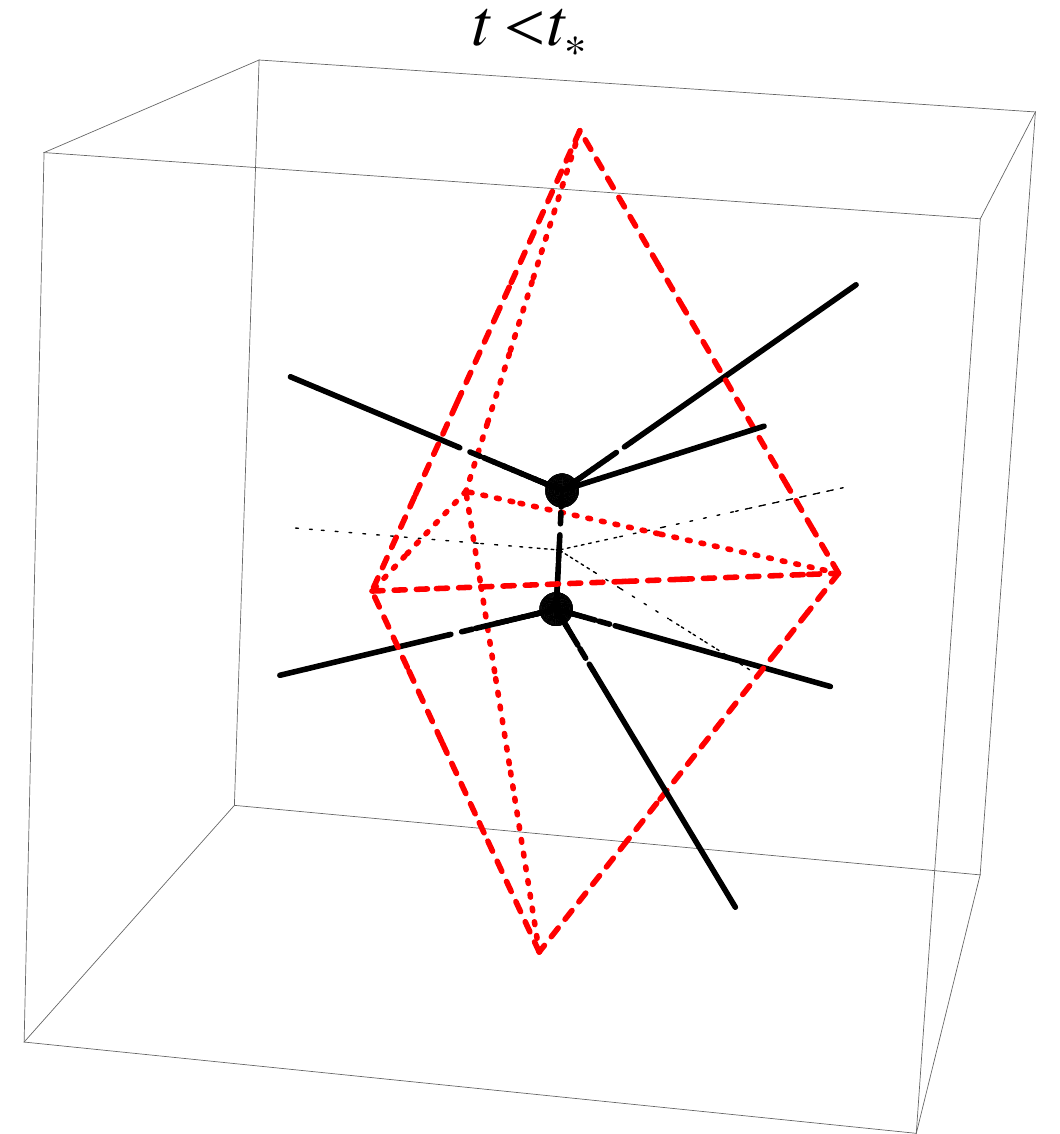}}
\epsfxsize=5 cm \epsfysize=5 cm {\epsfbox{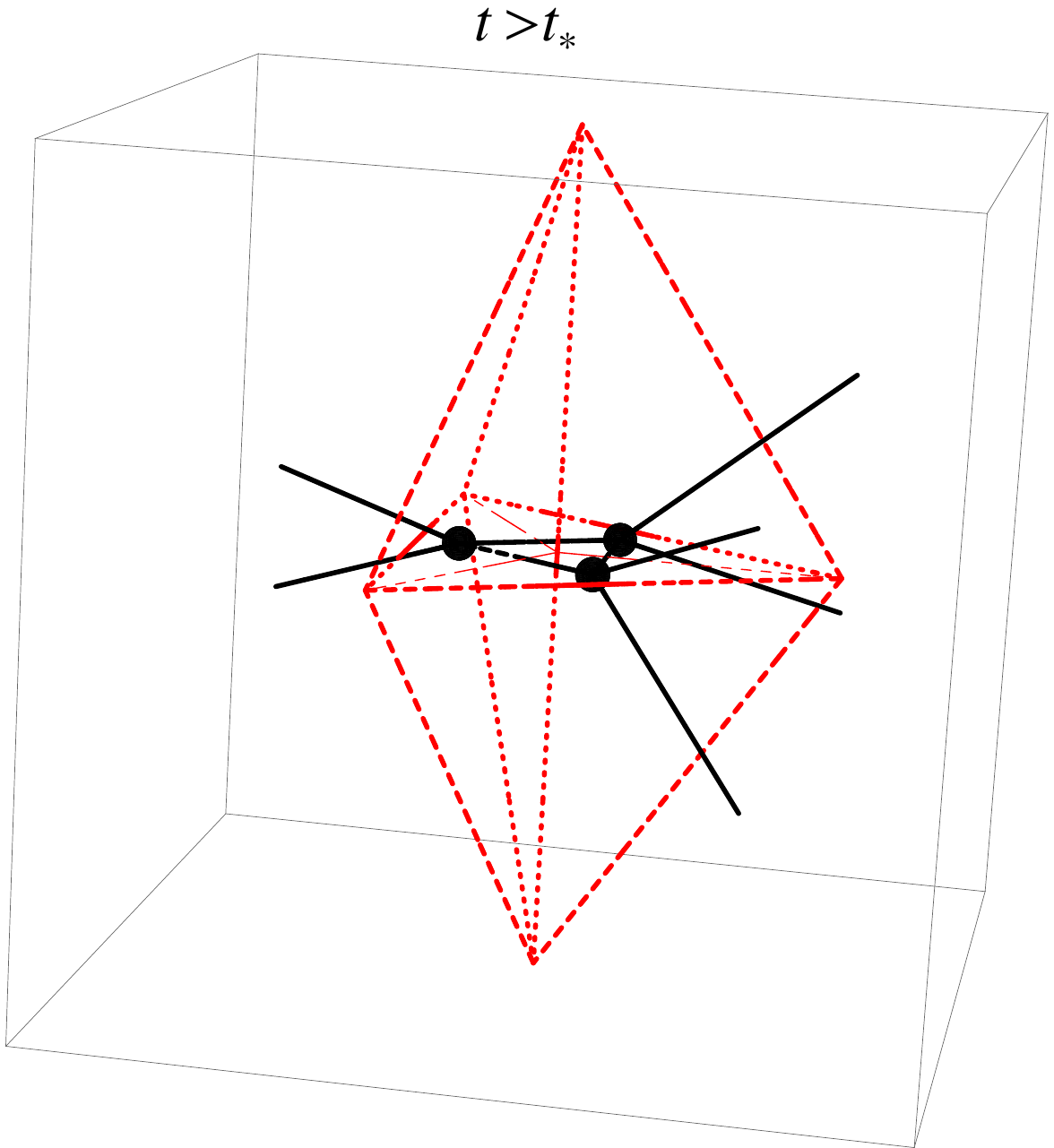}}
\end{center}
\caption{(Color online)
A collision of two shock nodes that gives rise to three new shock nodes, with a
redistribution of mass, in the three dimensional case. {\it Left panel}: at early times
we have two shock nodes (black dots), moving closer to the equatorial plane, which
contain the mass associated with the upper and lower tetrahedra (red dashed lines).
{\it Right panel}: after collision at $t_*$ we have three shock nodes moving apart
in the equatorial plane, while in Lagrangian space the triangular bipyramid built
by the two previous tetrahedra has been split into three new tetrahedra, with a
common edge set by the segment joining the upper an lower summits (vertical dashed
line).}
\label{figcoll_3d}
\end{figure*}

We now briefly discuss how the results illustrated in the previous sections in dimensions
$d=1$ and $d=2$ generalize to higher dimensions, and in particular to $d=3$.

As noticed in \cite{Gurbatov1991}, in three dimensions the Lagrangian-space partition is
made of tetrahedra (instead of triangles in $d=2$), while the Eulerian-space partition is
made of Voronoi cells with an arbitrary number of summits.
Then, the analysis developed in the previous sections shows that
the ``flip'' and ``three-merging'' events discussed in the two-dimensional case
in Sect.~\ref{Merging-fragmentation} generalize as the following rearrangements,

\begin{itemize}
\item $2\rightarrow 3$: two adjacent tetrahedra are reorganized into three new tetrahedra;
\item $3\rightarrow 2$: three adjacent tetrahedra are reorganized into two new tetrahedra;
\item $4\rightarrow 1$: merging of four adjacent tetrahedra into a single one.
\end{itemize}

These events naturally involve five vertices since, as seen from the paraboloid
construction (\ref{Paraboladef}), an Eulerian-space shock node is associated with
four Lagrangian-space vertices (where the paraboloid makes simultaneous first-contact
with $\psi_0(\bq)$), and rearrangements occur when a fifth summit encounters the
hyperplane defined by these four points. Higher-order events have a vanishingly small
probability.

The last event, $4\rightarrow 1$, which corresponds to the $3-$merging event observed
in $d=2$, occurs when a Lagrangian-space summit which is located within the tetrahedron
built by four neighboring summits is removed from the convex hull, as in the first step
shown in left panel in Fig.~\ref{figtime_2d}.
Indeed, before its removal, this interior summit leads to a splitting of the larger tetrahedron
into four distinct tetrahedra, associated with four shock nodes.
After the removal, only the embedding tetrahedron is left, which corresponds to a single
shock node that contains all the mass associated with the four previous shock nodes.

When the convex hull of the five Lagrangian vertices is not a tetrahedron
(i.e., no point is located within the tetrahedron formed by the other four summits),
one cannot build a single tetrahedron by complete merging, so that, as in the
two-dimensional events shown in middle panels of Fig.~\ref{figtime_2d},
one can only observe rearrangements of the matter distribution into new tetrahedra.
This leads to both events $2\rightarrow 3$ and $3\rightarrow 2$.

We illustrate the first case, $2\rightarrow 3$, in Fig.~\ref{figcoll_3d}. This corresponds
to Fig.~\ref{figcoll_2d}, associated with $2\rightarrow 2$ events in $d=2$.
We use the same notations and a similar symmetrical configuration. Thus,
we have three Lagrangian summits in the equatorial plane, $z=0$, which form an
equilateral triangle of side $\ell$ with the same value $\psi_0^{\rm equ}$, and
two symmetric
summits on the vertical axis at a larger distance $L$ with the same value
$\psi_0^{\rm u/d}>\psi_0^{\rm equ}$.
We again superpose the Eulerian-space Voronoi cells (with edges shown by the black
solid lines) and the Lagrangian-space tessellation built from elementary tetrahedra
(dashed red lines).
Then, at early times, using again the Hopf-Cole paraboloid solution (\ref{Paraboladef}), 
each Lagrangian summit is contained within its associated Voronoi cell and for $L$ 
large enough (or at small enough $t$) the central point $(0,0,0)$ ``sees'' the three
equatorial summits. This leads to the partition shown in the left panel with two
shock nodes.
As time increases, since $\psi_0^{\rm u/d}>\psi_0^{\rm equ}$ the Voronoi cells
associated with the
two upper and lower points expand and eventually make contact at the center,
at the time $t_*$ when the two shock nodes collide. Afterwards, these two
Voronoi cells have a common triangular facet in the equatorial plane, the cells associated
with the three equatorial summits being pushed outward, and we have three
outward-moving shock nodes.
In Lagrangian space, the two symmetric upper and lower tetrahedra shown in the left
panel, which merge into a unique volume at $t=t_*$, have been split into three new
tetrahedra, with a common edge given by the segment that joins the upper and lower
summits. 

As for the 2D case shown in Fig.~\ref{figcoll_2d}, it is interesting to note that the
``standard'' model would give a different behavior. Indeed, extending the discussion
of  Sect.~\ref{comparison}, we can see that using the standard continuity equation
we would obtain a single mass cluster that remains motionless in the center of the
figure. Thus, there is no fragmentation but the cluster leaves the shock nodes while
staying within the shock manifold. Note that this mass cluster is no longer located on
shock lines either, since it sits at the center of the horizontal triangular facet of the
shock manifold. This explicitly shows that within this ``standard'' model matter can
be distributed anywhere on the shock manifold, that is, not only on shock nodes
or shock lines, and that one needs to know the evolution of the system over all
previous times.

It is clear that the reverse event, $3\rightarrow 2$, can be obtained in a similar fashion,
by choosing the upper and lower summits close to the equatorial plane ($L<\ell$)
with $\psi_0^{\rm u/d}<\psi_0^{\rm equ}$.

These processes are naturally expected to generalize to higher dimensions, 
where the Lagrangian-space triangulation
is built from $(d+1)$-summit cells. Then, complete mergings are associated with
$(d+1)\rightarrow 1$ collisions in Eulerian space, while lower-order collisions,
$2\rightarrow d$, $3\rightarrow (d-1)$, ..., $d\rightarrow 2$, are associated with
Lagrangian-space rearrangements. The latter lead to a redistribution of matter,
so that particles which had coalesced into the same shock node at an earlier time
can be separated into distinct objects.
Moreover, such collisions also change the number of mass clusters in a very specific
manner, according to these rules.

\section{Conclusions, discussion}

The Burgers equation, and its Hopf-Cole solution, only determines the evolution of the velocity
field and to couple this to a transportation of matter one must add an equation for the evolution
of the density field. A natural choice would be to use
the ``standard'' continuity equation, as in \cite{Bogaevsky2004,Bogaevsky2006}, 
but it should then be integrated numerically over time. In the inviscid limit, it also
leads to behaviors that, at least in a cosmological context, are not necessarily realistic.



Here we rather explore an alternative choice, which we call the ``geometrical model'', that fully 
takes advantage of the Hopf-Cole solution to define the matter distribution from
a geometrical construction. It is based on Legendre transformations and convex hull
constructions, so that the matter distribution is associated with dual Eulerian and
Lagrangian space tessellations \cite{Gurbatov1991,Saichev1996}.
In regular regions and in the inviscid limit, both approaches coincide. The geometrical
model however corresponds to a non-standard continuity equation for the density field that affects
the mass behavior within the shock manifolds.

The peculiarities of this model are best revealed in the late time behavior of the density field
or for power-law spectra initial conditions. 
Then, in the inviscid limit that we investigate in this article, the matter is entirely contained in shock
nodes, which have gathered the matter that was initially in segments,
triangles, or tetrahedra, for respectively the $d=1$, $d=2$, or $d=3$ cases. In any
dimension, those objects form a partition of the Lagrangian space. 
%

The most striking result of these investigations is the mechanism with which 
the matter is rearranged in halos of growing mass as the system is evolving with time.
Indeed, for dimensions greater or equal to 2, these rearrangements follow a complex pattern
of successive ``flips'' and ``mergings''. 
In $d=2$, we explicitly show that these ``flips'' correspond in Lagrangian space to the
rearrangements of two triangles into two new triangles, and in Eulerian space to a
collision of two shock nodes giving rise to two new outward-moving shock nodes.
The ``merging'' events correspond in Lagrangian space to the merging of three triangles into
a single larger triangle, and in Eulerian space to the simultaneous merging of three halos
into a single node.
More complex patterns are expected for higher dimensional cases.

We have also described how the ``flips'' or ``fragmentation'' events, associated with
collisions that give rise to several new shock nodes, lead to a redistribution of momentum
over outgoing nodes. This implies that momentum is only conserved in a global
sense in $d \geq 2$ (whereas conservation of momentum also holds in a stronger
local sense in $d=1$, where there are only merging events).
Thus, in this regime the inviscid limit of the Burgers equation leads to a specific
set of collision rules (see note \footnote{They are at variance with familiar discrete systems, where in the limit of small particle
radii the dynamics is usually dominated by two-body collisions as higher-order
collisions are vanishingly rare.}), such as $\{2\rightarrow 2,3\rightarrow 1\}$ in 2D and
$\{2\rightarrow 3,3\rightarrow 2,4\rightarrow 1\}$ in 3D. In the present case, even though shock nodes are
infinitesimally thin, we have seen that one must take into account up to $(d+1)$-body
collisions in dimension $d$, which still occur at a finite rate. 

This peculiar behavior
is due to the geometrical construction that defines the matter distribution and that 
allows to integrate the equations of motion for both the velocity field
(using the standard Hopf-Cole solution of the Burgers equation) and the density
field (by definition of this geometrical construction). This is evidently a very convenient property that
has motivated this study. However, as shown here, this leads to a singular inviscid limit,
in the sense that particles that share the same location at a given time can separate at
later times -- such singular behaviors are associated with collisions between shock nodes -- so that it
does not lead to a genuine ``adhesion model''.
This result emphasizes the fact that the transportation of matter which can
be associated to the Burgers equation is a non-trivial process, whether one uses
the ``standard'' or the ``geometrical'' model, and that in the latter case one must be
aware of the peculiar merging and fragmentation events that take place as shock
nodes collide and reorganize the shock manifold.

\begin{acknowledgments}
The authors are especially thankful to Andrei Sobolevski and Uriel Frisch for their detailed
and fruitful comments on the first version of the manuscript. They helped clarify the
impact of our prescription for the continuity equation and pointed to previous literature
on the consequences of the alternative ``standard'' model.
This work is supported in part by the French Agence Nationale de la Recherche
under grant ANR-07-BLAN-0132 (BLAN07-1-212615).
\end{acknowledgments}

\bibliography{ref1}   
\end{document}